\documentclass[notitlepage,11pt]{article}

\usepackage{amssymb}
\usepackage{amsmath}
\usepackage{graphicx}

\input{tcilatex}

\begin{document}

\begin{center}
{\large Spin Glass Field Theory}

\smallskip

{\large with Replica Fourier Transforms}

\bigskip

I R Pimentel$^{1}$ and C De Dominicis$^{2}$

\bigskip

$^{1}$\textit{Departamento de F\'{\i}sica and CFTC, Faculdade de Ci\^{e}%
ncias,}

\textit{Universidade de Lisboa, Campo Grande 1749-016 Lisboa, Portugal}

$^{2}$\textit{CEA Saclay Orme des Merisiers, Inst Phys Theor,}

\textit{F-91191 Gif Sur Yvette, France}

\bigskip

\textit{E-mail:} irpimentel@fc.ul.pt, cirano.de-dominicis@cea.fr

\bigskip \bigskip

{\large Abstract }
\end{center}

\medskip

\noindent We develop a field theory for spin glasses using Replica Fourier
Transforms (RFT). We present the formalism for the case of replica symmetry
and the case of replica symmetry breaking on an ultrametric tree, with the
number of replicas $n$ and the number of replica symmetry breaking steps $R$
generic integers. We show how the RFT applied to the two-replica fields
allows to construct a new basis which block-diagonalizes the four-replica
mass-matrix, into the replicon, anomalous and longitudinal modes. The
eigenvalues are given in terms of the mass RFT and the propagators in the
RFT space are obtained by inversion of the block-diagonal matrix. The
formalism allows to express any $i$-replica vertex in the new RFT basis and
hence enables to perform a standard perturbation expansion. We apply the
formalism to calculate the contribution of the Gaussian fluctuations around
the Parisi solution for the free-energy of an Ising spin glass.

\bigskip \bigskip



\newpage

\noindent {\large 1. Introduction}

\bigskip

Spin glasses are disordered magnetic systems with frustration [1-4].These
systems exhibit a \textit{freezing} transition to a low temperature phase
with nontrivial properties. Although spin glasses have been studied for over
three decades there is still no consensus on the nature of the glassy phase.
Two different pictures have been proposed for the spin glass. One
corresponds to the Parisi solution [5] of the infinite-range
Sherrington-Kirkpatrick (SK) model [6], which represents the mean field
theory for spin glasses and predicts a glassy phase described by an infinite
number of pure states, organized in an ultrametric structure. The other one
is the "droplet" model [7-9], which claims that in the experimentally
relevant short-range spin glasses the glassy phase is described by only two
pure states, related by a global inversion of the spins. The first picture
is set within a replica field theory and results from replica symmetry
breaking, while the second picture is based on a scaling theory and
corresponds to no replica symmetry breaking. An important step for the
understanding of spin glasses, lies in the investigation of how the
fluctuations, associated into the finite-range interactions modify the
mean-field picture.

Edwards and Anderson [10] introduced a model for short-range spin glasses
and used the replica method to perform the average over quenched disorder. A
field theory is built for the spin glass with the free energy being written
as a functional of replica fields $Q_{i}^{ab}$ (where $a=1,\ldots ,n$ is a
replica index), which represent the spin glass order parameter. A
perturbation expansion around the mean-field solution, which corresponds to
the infinite-range or infinite-dimensional (i.e., spin coordination number $%
z\rightarrow \infty $) model is constructed by separating the field $%
Q_{i}^{ab}$ into its mean field value $Q^{ab}$ and fluctuations $\phi
_{i}^{ab}$ around it. The mean field value of the order parameter $Q^{ab}$
is provided by the stationarity condition of the free energy and the
stability of the solution is determined by the analysis of the Hessian or
mass-matrix $M^{ab,cd}$ of the fluctuations, that is by the evaluation of
the eigenvalues or, in other terms, the diagonalization of the matrix. In
turn, to calculate physical properties one needs the propagators $G^{ab,cd}$
of the fluctuations, the bare propagators being given by the inverse of the
mass-matrix. The replica dependence of $Q^{ab}$, which reflects the
structure of the order parameter, naturally determines the form of the
mass-matrix, and also the form of the bare propagators and the interaction
vertices of the fluctuations for higher order calculations in the
perturbation expansion.

In mean field theory it is found that a phase transition occurs at a
critical temperature, from a high-temperature phase with replica symmetry
(RS) to a low-temperature phase with replica symmetry breaking (RSB). The
stability of the RS solution was studied by de Almeida and Thouless [11],
who provided the eigenvalues, their multiplicities, and the eigenvectors of
the mass-matrix. They found three different sets of modes, later called
replicon, anomalous and longitudinal [12], with the longitudinal and
anomalous eigenvalues becoming equal for $n=0$. In the presence of a
magnetic field the instability of the RS solution against RSB occurs along a
line in the temperature-field plane, the Almeida-Thouless (AT) line. In zero
field all the modes become critical at the transition temperature, while in
nonzero field only the replicon mode becomes critical at the AT line. The
propagators for the RS theory were obtained by Bray and Moore [12] and Pytte
and Rudnick [13]. The RSB ansatz proposed by Parisi for the spin glass,
which turns out to be the exact solution for the SK model [14], represents
many states in a hierarchical organization that is described by an
ultrametric tree. The study of the stability of the Parisi RSB solution is a
nontrivial task that was carried out by De Dominicis and Kondor [15]. They
found that the eigenvalues of the mass-matrix form two continuous bands, for 
$n=0$, corresponding to replicon and longitudinal-anomalous modes, the lower
band, associated to the replicon, being bounded below by zero. Fairly
involving computations performed by Kondor and De Dominicis [16] and De
Dominicis and Kondor [17] led to results for the multiplicities and
propagators of the RSB theory. See De Dominicis \textit{et al} [18] for a
review on the spin glass field theory with RSB. The high complexity of the
theory has however inhibited the study of the glassy phase.

A fundamental aspect for the study of spin glasses is the diagonalization
and inversion of the ultrametric four-replica mass-matrix, which turns out
to be a rather difficult problem. Temesv\'{a}ri \textit{et al} [19] and De
Dominicis \textit{et al} [20] provided results for the block-diagonalization
and inversion of the mass-matrix in direct replica space. The block
diagonalized form is a consequence of the ultrametric symmetry of the matrix
which reflects the residual symmetry of the problem, after the Parisi
breaking of replica symmetry. De Dominicis \textit{et al} [21] later used
the concept of Replica Fourier Transform to block-diagonalize and invert the
mass-matrix, clearly showing the advantage of this method.

In this article we develop a field theory for spin glasses using Replica
Fourier Transforms (RFT). We consider both the case of a replica symmetric
theory where the simple RFT is used and the case of replica symmetry
breaking where the RFT is defined on a tree. We show how the RFT applied to
the two-replica fields leads to a new basis which block-diagonalizes the
four-replica mass-matrix, into three sets of modes, replicon, anomalous and
longitudinal. The eigenvalues of the replicon, anomalous and longitudinal
modes are then given in terms of the RFT of the mass-matrix. The
corresponding multiplicities and eigenvectors are provided. The propagators
in the RFT space are then readily obtained by inversion of the
block-diagonal mass-matrix. The formalism allows to express any $i$-replica
vertex in the new RFT basis, and hence enables to perform a standard
perturbation expansion. We keep the number $n$ of replicas a positive
integer, the limit $n\rightarrow 0$ of the replica method can be taken at
the very end, on the final results. The number of replica symmetry breaking
steps $R$ is also considered a generic integer, hence our results apply
either in a situation where only a single RSB step is needed, or in the case
of full RSB $R\rightarrow \infty $ proposed by Parisi. We show that many
fundamental results for the study of spin glasses, can be simply derived
within the RFT formalism.

The outline of this article is as follows. In Section 2 we present the field
theory for an Ising spin glass in direct replica space. In Section 3 we
develop the RFT formalism for the replica symmetric case $R=0$. In Section 4
we generalize the RFT formalism to the case of replica symmetry breaking $%
R\neq 0$. In Section 5 we apply the formalism to calculate the contribution
of the Gaussian fluctuations around the Parisi solution for the free-energy
of an Ising spin glass, which illustrates the physical relevance of the
results presented. Section 6 concludes the article with an overview of the
work.

\bigskip \bigskip

\noindent {\large 2. Spin Glass Model}

\bigskip

We consider an Ising spin glass in a uniform magnetic field $H$, described
by the Edwards-Anderson model%
\begin{equation}
\mathcal{H}=-\underset{\left\langle ij\right\rangle }{\sum }%
J_{ij}S_{i}S_{j}-H\underset{i}{\sum }S_{i}  \label{1}
\end{equation}

\noindent for $N$ spins, $S_{i}=\pm 1$, located on a regular $d$-dimensional
lattice, where the bonds $J_{ij}$, which couple nearest-neighbor spins only,
are independent random variables with a Gaussian distribution, characterized
by zero mean and variance $\Delta ^{2}=J^{2}/z$, $z=2d$ being the
coordination number. The summations are over pairs $\left\langle
ij\right\rangle $ of distinct sites on the lattice and over the lattice
sites $i$.

The free energy averaged over the quenched disorder is given, via the
replica method, by%
\begin{equation}
\overline{F}=-\frac{1}{\beta }\overline{\ln Z}=-\frac{1}{\beta }\underset{%
n\rightarrow 0}{\lim }\frac{\overline{Z^{n}}-1}{n}  \label{2}
\end{equation}%
where $Z$ is the partition function and $\beta =1/k_{B}T$.

Taking the average of $n$ replicas of the partition function $\overline{Z^{n}%
}$, with $n$ integer, followed by a Hubbard-Stratonovich transformation, to
decouple a four-spin term, leads to%
\begin{equation}
\overline{Z^{n}}=\int \underset{(ab);i}{\prod }\frac{dQ_{i}^{ab}}{\sqrt{2\pi 
}}\exp \left\{ -\mathcal{L}\left[ Q_{i}^{ab}\right] \right\}  \label{3}
\end{equation}

\noindent with%
\begin{eqnarray}
\mathcal{L}\left[ Q_{i}^{ab}\right] &=&-\frac{Nn\left( \beta J\right) ^{2}}{4%
}+\frac{z\left( \beta J\right) ^{2}}{2}\underset{i,j}{\sum }\underset{(ab)}{%
\sum }Q_{i}^{ab}(K^{-1})_{ij}Q_{j}^{ab}  \label{4} \\
&&-\underset{i}{\sum }\ln \underset{\left\{ S_{i}^{a}\right\} }{\text{Tr}}%
\exp \left\{ \left( \beta J\right) ^{2}\underset{\left( ab\right) }{\sum }%
Q_{i}^{ab}S_{i}^{a}S_{i}^{b}+\beta H\underset{a}{\sum }S_{i}^{a}\right\} 
\notag
\end{eqnarray}

\noindent where $K_{ij}=1$ for nearest neighbor sites and $0$ otherwise, and 
$S_{i}^{a}$ are spins with replica index $a=1,\ldots ,n$. The fields $%
Q_{i}^{ab}$ are defined on an $n(n-1)/2$-dimensional replica space of pairs $%
(ab)$ of distinct replicas, since $Q_{i}^{ab}=Q_{i}^{ba}$ and $Q_{i}^{aa}=0$.

In order to construct a perturbation expansion around the mean-field
solution, one separates the field $Q_{i}^{ab}$ into 
\begin{equation}
Q_{i}^{ab}=Q^{ab}+\phi _{i}^{ab}  \label{5}
\end{equation}

\noindent where $Q^{ab}$ represents the mean field order parameter\ and$\
\phi _{i}^{ab}$ are fluctuations around it. The Lagrangian $\mathcal{L}$ is
then given by%
\begin{equation}
\mathcal{L}=\mathcal{L}^{(0)}+\mathcal{L}^{(1)}+\mathcal{L}^{(2)}+\ldots
\label{6}
\end{equation}%
where, after Fourier transform into momenta space, one has, for
contributions up to quadratic order in the fluctuations,%
\begin{eqnarray}
\mathcal{L}^{(0)} &=&-\frac{Nn\left( \beta J\right) ^{2}}{4}+\frac{N(\beta
J)^{2}}{2}\underset{(ab)}{\sum }(Q^{ab})^{2}  \label{7} \\
&&-N\ln \underset{\left\{ S^{a}\right\} }{\text{Tr}}\exp \left\{ (\beta
J)^{2}\underset{(ab)}{\sum }Q^{ab}S^{a}S^{b}+\beta H\underset{a}{\sum }%
S^{a}\right\}  \notag
\end{eqnarray}

\begin{equation}
\mathcal{L}^{(1)}=\sqrt{N}(\beta J)^{2}\underset{(ab)}{\sum }\left[
Q^{ab}-\left\langle S^{a}S^{b}\right\rangle \right] \phi _{\mathbf{p}=%
\mathbf{0}}^{ab}  \label{8}
\end{equation}%
\begin{equation}
\mathcal{L}^{(2)}=\frac{1}{2}\underset{(ab)(cd)}{\sum }\underset{\mathbf{p}}{%
\sum }\phi _{\mathbf{p}}^{ab}M^{ab,cd}(\mathbf{p})\phi _{-\mathbf{p}}^{cd}
\label{9}
\end{equation}

\noindent with%
\begin{eqnarray}
M^{ab,cd}(\mathbf{p}) &=&\mathbf{p}^{2}\delta _{ab,cd}^{Kr}+z\left[ \delta
_{ab,cd}^{Kr}-(\beta J)^{2}\left( \left\langle
S^{a}S^{b}S^{c}S^{d}\right\rangle \right. \right.  \label{10} \\
&&\qquad \qquad \qquad -\left. \left. \left\langle S^{a}S^{b}\right\rangle
\left\langle S^{c}S^{d}\right\rangle \right) \right]  \notag
\end{eqnarray}

\noindent where $S^{a}=S_{i}^{a}$ and the expectation value $\left\langle
\cdots \right\rangle $ is calculated with the normalized weight $\zeta (S)/$%
Tr$\zeta (S)$, where%
\begin{equation}
\zeta (S)=\exp \left\{ (\beta J)^{2}\underset{(ab)}{\sum }%
Q^{ab}S^{a}S^{b}+\beta H\sum_{a}S^{a}\right\} .  \label{11}
\end{equation}

\noindent In (9), the sum in momenta is confined to the range $0<\left\vert 
\mathbf{p}\right\vert <\Lambda $, with a cutoff $\Lambda \simeq 1$, the
mass-matrix $M^{ab,cd}(\mathbf{p})$ is expanded for small $\mathbf{p}$,
keeping only the terms up to second order, and the fields are rescaled $%
\left[ \phi \left( \beta J/\sqrt{z}\right) \rightarrow \phi \right] $ to
allow to write the coefficient of the momentum equal to unity.

The mean-field value of the order parameter $Q^{ab}$ is determined by the
stationarity condition $\mathcal{L}^{(1)}=0$, which from (8) gives%
\begin{equation}
Q^{ab}=\left\langle S^{a}S^{b}\right\rangle .  \label{12}
\end{equation}

\noindent Hence, $Q^{ab}$ represents the spin overlap between replicas $a$
and $b$. Considering \ a replica symmetric (RS) solution, $Q^{ab}=Q$, (12)
leads to the following results. In zero magnetic field, $H=0$, there is a
phase transition at a critical temperature $T_{c}=J/k_{B}$: $Q=0$ for $T\geq
T_{c}$, while $Q\neq 0$ for $T<T_{c}$. However, the RS solution turns out to
be unstable in the low-temperature phase, and replica symmetry breaking
(RSB) is required. In a nonzero magnetic field, $H\neq 0$, there is a phase
transition along a line in the field-temperature plane, the AT line, which
in the region of small fields $H$, and near the zero-field critical
temperature $T_{c}$, is given by $(H/J)^{2}=(4/3)(1-T/T_{c})^{3}$: above the
AT\ line a RS solution $Q\neq 0$ is stable, while below the AT line the RS
solution becomes unstable and RSB is required.

The normal modes of the fluctuations of the order parameter are obtained by
re-writing $\mathcal{L}^{(2)}$, (9), in a diagonal form. The eigenvalues of
the matrix $M^{ab,cd}$ are then provided, and the propagators can be easily
obtained by inversion of the diagonalized matrix.

\bigskip \bigskip \bigskip

\noindent {\large 3. Replica Symmetric Ansatz }

\bigskip

Here we consider that the mean-field order parameter is replica symmetric%
\begin{equation}
Q^{ab}=Q,\quad a\neq b.  \label{13}
\end{equation}

\noindent In this case, there are three distinct masses%
\begin{eqnarray}
M^{ab,ab} &=&M_{11}  \notag \\
M^{ab,ac} &=&M^{ab,bc}=M_{10}  \label{14} \\
M^{ab,cd} &=&M_{00}.  \notag
\end{eqnarray}

\noindent The Lagrangian term of the fluctuations $\mathcal{L}^{(2)}$, (9),
then takes the form%
\begin{equation}
\mathcal{L}^{(2)}\mathcal{=}\frac{1}{2}\left\{ M_{11}\underset{(ab)}{\sum }%
\phi _{ab}^{2}+M_{10}\underset{(abc)}{\sum }\left( \phi _{ab}\phi _{ac}+\phi
_{ab}\phi _{bc}\right) +M_{00}\underset{(abcd)}{\sum }\phi _{ab}\phi
_{cd}\right\}  \label{15}
\end{equation}

\noindent where the\ dependence on momentum $\mathbf{p}$ is implicit and the
sums are restricted to distinct replicas.

Writing $\mathcal{L}^{(2)}$ in terms of sums over unrestricted replicas, one
obtains%
\begin{eqnarray}
\mathcal{L}^{(2)} &=&\frac{1}{4}\left\{ \left( M_{11}-2M_{10}+M_{00}\right) 
\underset{a,b}{\sum }\phi _{ab}^{2}\right.  \label{16} \\
&&+\left( M_{10}-M_{00}\right) \underset{a,b,c}{\sum }(\phi _{ab}\phi
_{ac}+\phi _{ab}\phi _{bc})\left. +\frac{1}{2}M_{00}\underset{a,b,c,d}{\sum }%
\phi _{ab}\phi _{cd}\right\}  \notag
\end{eqnarray}%
with the field constraints%
\begin{equation}
\phi _{aa}=0,\quad a=1,\ldots ,n.  \label{17}
\end{equation}

The RFT for a field with a single replica index, and its inverse
transformation, are defined as%
\begin{eqnarray}
\phi _{\hat{a}} &=&\frac{1}{\sqrt{n}}\underset{a}{\sum }e^{-\frac{2\pi i}{n}a%
\hat{a}}\phi _{a}  \label{18} \\
\phi _{a} &=&\frac{1}{\sqrt{n}}\underset{\hat{a}}{\sum }e^{\frac{2\pi i}{n}a%
\hat{a}}\phi _{\hat{a}}  \notag
\end{eqnarray}

\noindent with $a=1,\ldots ,n,$ $\hat{a}=0,\ldots ,n-1,$ and $a,$ $\hat{a}$
considered $\func{mod}(n)$. One has the relation%
\begin{equation}
\underset{a}{\sum }e^{\frac{2\pi i}{n}a\hat{a}}=n\delta _{\hat{a},\hat{0}}
\label{19}
\end{equation}

\noindent (using $\hat{0}$ when $\hat{a}=0$). From (18), it follows that $%
\phi _{\hat{a}}^{\ast }=$ $\phi _{-\hat{a}}$. For the two-replica fields we
then have%
\begin{eqnarray}
\phi _{ab} &=&\frac{1}{n}\underset{\hat{a}\hat{b}}{\sum }e^{\frac{2\pi i}{n}%
(a\hat{a}+b\hat{b})}\phi _{\hat{a}\hat{b}}  \label{20} \\
\phi _{\hat{a}\hat{b}} &=&\frac{1}{n}\underset{ab}{\sum }e^{-\frac{2\pi i}{n}%
(a\hat{a}+b\hat{b})}\phi _{ab}  \notag
\end{eqnarray}

\noindent with the symmetry $\phi _{\hat{a}\hat{b}}=\phi _{\hat{b}\hat{a}}$
resulting from $\phi _{ab}=\phi _{ba}$. The fields can be written as $\phi _{%
\hat{a}\hat{b}}=\phi _{\hat{a},\hat{t}-\hat{a}}$, where $\hat{t}=\hat{a}+%
\hat{b}$. For $\hat{t}=0$ the fields are real, while for $\hat{t}\neq 0$
they are complex, with $\phi _{\hat{a},\hat{t}-\hat{a}}^{\ast }=\phi _{-\hat{%
a},-\hat{t}+\hat{a}}$.

After RFT the Lagrangian $\mathcal{L}^{(2)}$ becomes,%
\begin{eqnarray}
\mathcal{L}^{(2)} &=&\frac{1}{4}\left\{ \left( M_{11}-2M_{10}+M_{00}\right) 
\underset{\hat{t},\hat{a}}{\sum }\left\vert \phi _{\hat{a},\hat{t}-\hat{a}%
}\right\vert ^{2}\right.  \label{21} \\
&&+\left. 2n\left( M_{10}-M_{00}\right) \sum_{\hat{t}}\left\vert \phi _{\hat{%
0},\hat{t}}\right\vert ^{2}+\frac{n^{2}}{2}M_{00}\left\vert \phi _{\hat{0},%
\hat{0}}\right\vert ^{2}\right\}  \notag
\end{eqnarray}

\noindent with the field constraints in (17) expressed as%
\begin{equation}
\underset{\hat{a}=0}{\sum^{n-1}}\phi _{\hat{a},\hat{t}-\hat{a}}=0,\qquad 
\hat{t}=0,\ldots ,n-1  \label{22}
\end{equation}%
which follows from taking the RFT of $\phi _{aa}$ over the index $a$.

Separating in (21) the fields with indices $\hat{0}$, one obtains%
\begin{eqnarray}
\mathcal{L}^{(2)} &=&\frac{1}{4}\left\{ \left( M_{11}-2M_{10}+M_{00}\right)
\left( \underset{\hat{a}^{\prime }}{\sum }\left\vert \phi _{\hat{a}^{\prime
},-\hat{a}^{\prime }}\right\vert ^{2}+\underset{\hat{t}^{\prime },\hat{a}%
^{\prime \prime }}{\sum }\left\vert \phi _{\hat{a}^{\prime \prime },\hat{t}%
^{\prime }-\hat{a}^{\prime \prime }}\right\vert ^{2}\right) \right.  \notag
\\
&&+2\left( M_{11}+(n-2)M_{10}-(n-1)M_{00}\right) \sum_{\hat{t}^{\prime
}}\left\vert \phi _{\hat{0},\hat{t}^{\prime }}\right\vert ^{2}  \label{23} \\
&&+\left. \left( M_{11}+2\left( n-1\right) M_{10}+\left( 1-2n+\frac{n^{2}}{2}%
\right) M_{00}\right) \left\vert \phi _{\hat{0},\hat{0}}\right\vert
^{2}\right\}  \notag
\end{eqnarray}%
where $\hat{t}^{\prime }\neq \hat{0}$, $\hat{a}^{\prime }\neq \hat{0}$ and $%
\hat{a}^{\prime \prime }\neq \hat{0},\hat{t}^{\prime }$.

Now, we define the new fields%
\begin{equation}
\Phi _{\hat{a}^{\prime },-\hat{a}^{\prime }}=\phi _{\hat{a}^{\prime },-\hat{a%
}^{\prime }}+\frac{1}{n-1}\phi _{\hat{0},\hat{0}}  \label{24}
\end{equation}

\noindent and%
\begin{equation}
\Phi _{\hat{a}^{\prime \prime },\hat{t}^{\prime }-\hat{a}^{\prime \prime
}}=\phi _{\hat{a}^{\prime \prime },\hat{t}^{\prime }-\hat{a}^{\prime \prime
}}+\frac{1}{n-2}\left( \phi _{\hat{0},\hat{t}^{\prime }}+\phi _{\hat{t}%
^{\prime },\hat{0}}\right)  \label{25}
\end{equation}

\noindent which, from (22), have the constraints%
\begin{equation}
\underset{\hat{a}^{\prime }}{\sum }\Phi _{\hat{a}^{\prime },-\hat{a}^{\prime
}}=0,  \label{26}
\end{equation}

\noindent and%
\begin{equation}
\underset{\hat{a}^{\prime \prime }}{\sum }\Phi _{\hat{a}^{\prime \prime },%
\hat{t}^{\prime }-\hat{a}^{\prime \prime }}=0.  \label{27}
\end{equation}

\noindent Then, by introducing the new fields in (23), one obtains the
Lagrangian $\mathcal{L}^{(2)}$ in the diagonal form,%
\begin{eqnarray}
\mathcal{L}^{(2)} &=&\frac{1}{2}\left\{ M_{R}\underset{\hat{a}^{\prime }}{%
\sum }\left\vert ^{R}\Phi _{\hat{a}^{\prime },-\hat{a}^{\prime }}\right\vert
^{2}+M_{R}\underset{\hat{t}^{\prime },\hat{a}^{\prime \prime }}{\sum }%
\left\vert ^{R}\Phi _{\hat{a}^{\prime \prime },\hat{t}^{\prime }-\hat{a}%
^{\prime \prime }}\right\vert ^{2}\right.  \label{28} \\
&&\left. +M_{A}\sum_{\hat{t}^{\prime }}\left\vert ^{A}\phi _{\hat{0},\hat{t}%
^{\prime }}\right\vert ^{2}+M_{L}\left\vert ^{L}\phi _{\hat{0},\hat{0}%
}\right\vert ^{2}\right\}  \notag
\end{eqnarray}

\noindent where%
\begin{eqnarray}
M_{R} &=&M_{11}-2M_{10}+M_{00}  \notag \\
M_{A} &=&M_{11}+(n-4)M_{10}-(n-3)M_{00}  \label{29} \\
M_{L} &=&M_{11}+2(n-2)M_{10}+\frac{1}{2}\left( n-2\right) (n-3)M_{00}  \notag
\end{eqnarray}

\noindent and 
\begin{eqnarray}
^{R}\Phi _{\hat{a}^{\prime },-\hat{a}^{\prime }} &=&\frac{1}{\sqrt{2}}\Phi _{%
\hat{a}^{\prime },-\hat{a}^{\prime }};\quad ^{R}\Phi _{\hat{a}^{\prime
\prime },\hat{t}^{\prime }-\hat{a}^{\prime \prime }}=\frac{1}{\sqrt{2}}\Phi
_{\hat{a}^{\prime \prime },\hat{t}^{\prime }-\hat{a}^{\prime \prime }}
\label{30} \\
^{A}\phi _{\hat{0},\hat{t}^{\prime }} &=&\sqrt{\frac{n}{(n-2)}}\phi _{\hat{0}%
,\hat{t}^{\prime }};\quad ^{L}\phi _{\hat{0},\hat{0}}=\sqrt{\frac{n}{2(n-1)}}%
\phi _{\hat{0},\hat{0}}  \notag
\end{eqnarray}%
with the constraints,%
\begin{equation}
\underset{\hat{a}^{\prime }}{\sum }{}^{R}\Phi _{\hat{a}^{\prime },-\hat{a}%
^{\prime }}=0;\quad \underset{\hat{a}^{\prime \prime }}{\sum }{}^{R}\Phi _{%
\hat{a}^{\prime \prime },\hat{t}^{\prime }-\hat{a}^{\prime \prime }}=0.
\label{31}
\end{equation}

\noindent The fields $^{L}\phi _{\hat{0},\hat{0}}$, $^{A}\phi _{\hat{0},\hat{%
t}^{\prime }}$, $^{R}\Phi _{\hat{a}^{\prime },-\hat{a}^{\prime }}$ and $%
^{R}\Phi _{\hat{a}^{\prime \prime },\hat{t}^{\prime }-\hat{a}^{\prime \prime
}}$ are symmetrized: $\phi _{\hat{0},\hat{t}}=\frac{1}{2}(\phi _{\hat{0},%
\hat{t}}+\phi _{\hat{t},\hat{0}})$ \ and $\;\Phi _{\hat{a}^{\prime \prime },%
\hat{t}-\hat{a}^{\prime \prime }}=\frac{1}{2}(\Phi _{\hat{a}^{\prime \prime
},\hat{t}-\hat{a}^{\prime \prime }}+\Phi _{\hat{t}-\hat{a}^{\prime \prime },%
\hat{a}^{\prime \prime }})$,\ for $\hat{t}=0$\ and$\;\hat{t}\neq 0$; \ the
fields\ $\ ^{L}\phi _{\hat{0},\hat{0}}\;\ $and $\ ^{A}\phi _{\hat{0},\hat{t}%
^{\prime }}$ \ are also normalized, the normalization of\ $\;^{R}\Phi _{\hat{%
a}^{\prime },-\hat{a}^{\prime }}$\ \ is $\;N_{1}=\sqrt{\left( n-3\right)
/(2\left( n-1\right) )}$ \ and that of $\;^{R}\Phi _{\hat{a}^{\prime \prime
},\hat{t}^{\prime }-\hat{a}^{\prime \prime }}\;\;$is $\;N_{2}=\sqrt{(\delta
_{\hat{t}^{\prime },2\hat{a}^{\prime \prime }}+\left( n-4\right) /\left(
n-2\right) )/2}$.

One can see from (28) that the fluctuation space is divided into three
sectors, which we identify as the replicon (R) with eigenvalue $M_{R}$, the
anomalous (A) with eigenvalue $M_{A}$, and the longitudinal (L) with
eigenvalue $M_{L}$. The degeneracies of the eigenvalues are given by the
multiplicities of the fields, $\mu _{1}=(n-3)/2$ for $^{R}\Phi _{\hat{a}%
^{\prime },-\hat{a}^{\prime }}$ and $\mu _{2}=(n-1)(n-3)/2$ for $^{R}\Phi _{%
\hat{a}^{\prime \prime },\hat{t}^{\prime }-\hat{a}^{\prime \prime }}$ leads
to $\mu _{R}=\mu _{1}+\mu _{2}=n(n-3)/2$ for the replicon, $\mu _{A}=(n-1)$
for the anomalous and $\mu _{L}=1$ for the longitudinal, so that the total
number of modes is recovered $\mu _{R}+\mu _{A}+\mu _{L}=n(n-1)/2$.

We note that the replicon, anomalous and longitudinal masses are given in
terms of the RFT of the original masses as%
\begin{eqnarray}
M_{R} &=&\hat{M}_{11}  \notag \\
M_{A} &=&\hat{M}_{11}+\frac{1}{4}(n-2)\hat{M}_{10}  \label{32} \\
M_{L} &=&\hat{M}_{11}+\frac{1}{2}(n-1)\hat{M}_{00}  \notag
\end{eqnarray}

\noindent where the RFT of the original masses are defined as [21]%
\begin{eqnarray}
\hat{M}_{11} &=&M_{11}-2M_{10}+M_{00}  \notag \\
\hat{M}_{10} &=&4(M_{10}-M_{00})  \label{33} \\
\hat{M}_{00} &=&4(M_{10}-M_{00})+nM_{00}.  \notag
\end{eqnarray}

The propagators for the longitudinal, anomalous and replicon modes can be
easily obtained from (28) and are given by%
\begin{equation}
^{L}G_{\hat{0};\hat{0}}=\left\langle ^{L}\phi _{\hat{0},\hat{0}}{}^{L}\phi _{%
\hat{0},\hat{0}}\right\rangle =\frac{1}{M_{L}}  \label{34}
\end{equation}%
\begin{equation}
^{A}G_{\hat{t}^{\prime };\hat{s}^{\prime }}=\left\langle ^{A}\phi _{\hat{0},%
\hat{t}^{\prime }}{}^{A}\phi _{\hat{0},\hat{s}^{\prime }}^{\ast
}\right\rangle =\delta _{\hat{t}^{\prime },\hat{s}^{\prime }}\frac{1}{M_{A}}
\label{35}
\end{equation}%
\begin{eqnarray}
^{R}G_{\hat{a}^{\prime };\hat{b}^{\prime }} &=&\left\langle ^{R}\Phi _{\hat{a%
}^{\prime },-\hat{a}^{\prime }}{}^{R}\Phi _{\hat{b}^{\prime },-\hat{b}%
^{\prime }}\right\rangle  \label{36} \\
&=&\frac{1}{2}\left[ \left( \delta _{\hat{a}^{\prime },\hat{b}^{\prime
}}+\delta _{\hat{a}^{\prime },-\hat{b}^{\prime }}\right) -\frac{2}{n-1}%
\right] \frac{1}{M_{R}}  \notag
\end{eqnarray}%
\begin{eqnarray}
^{R}G_{\hat{a}^{\prime \prime },\hat{t}^{\prime };\hat{b}^{\prime \prime },%
\hat{s}^{\prime }} &=&\left\langle ^{R}\Phi _{\hat{a}^{\prime \prime },\hat{t%
}^{\prime }-\hat{a}^{\prime \prime }}{}^{R}\Phi _{\hat{b}^{\prime \prime },%
\hat{s}^{\prime }-\hat{b}^{\prime \prime }}^{\ast }\right\rangle  \label{37}
\\
&=&\delta _{\hat{t}^{\prime },\hat{s}^{\prime }}\frac{1}{2}\left[ \left(
\delta _{\hat{a}^{\prime \prime },\hat{b}^{\prime \prime }}+\delta _{\hat{a}%
^{\prime \prime },\hat{s}^{\prime }-\hat{b}^{\prime \prime }}\right) -\frac{2%
}{n-2}\right] \frac{1}{M_{R}}.  \notag
\end{eqnarray}

\noindent We remark that the propagators for the replicon, (36) and (37),
are not completely diagonalized because of the constraints in (31).

The propagators in the direct replica space can be easily obtained, in terms
of their RFT expression, e.g.,%
\begin{eqnarray}
G^{ab,ab} &=&G_{11}=\left\langle \phi _{ab}\phi _{ab}\right\rangle =\frac{1}{%
n(n-1)}\underset{ab}{\sum }\left\langle \phi _{ab}\phi _{ab}\right\rangle 
\notag \\
&=&\frac{1}{n(n-1)}\underset{\hat{t},\hat{a}=0}{\sum }\left\langle \phi _{%
\hat{a},\hat{t}-\hat{a}}\phi _{\hat{a},\hat{t}-\hat{a}}^{\ast }\right\rangle
\label{38} \\
&=&\frac{2}{n(n-1)}\left[ ^{L}G_{\hat{0},\hat{0}}+\underset{\hat{t}^{\prime }%
}{\sum }{}^{A}G_{\hat{t}^{\prime },\hat{t}^{\prime }}+\underset{\hat{a}%
^{\prime }}{\sum }{}^{R}G_{\hat{a}^{\prime },\hat{a}^{\prime }}+\underset{%
\hat{t}^{\prime },\hat{a}^{\prime \prime }}{\sum }{}^{R}G_{\hat{a}^{\prime
\prime },\hat{t}^{\prime };\hat{a}^{\prime \prime },\hat{t}^{\prime }}\right]
.  \notag
\end{eqnarray}

\bigskip \bigskip

\noindent {\large 4. Replica Symmetry Breaking: Parisi's Ansatz }

\bigskip

The replica symmetry breaking ansatz proposed by Parisi for the mean-field
order parameter can be described as follows. Consider $Q^{ab}$ as a
symmetric $n\times n$ matrix with zeros on its diagonal. One starts with the
replica symmetric form, in which all the off-diagonal elements have the same
value, $Q_{0}$. We then divide the $n\times n$ matrix into blocks of size $%
p_{1}\times p_{1}$, and in the diagonal blocks replace $Q_{0}$ by $Q_{1}$,
leaving $Q_{0}$ in the off-diagonal blocks. Each of the $p_{1}\times p_{1}$
blocks on the diagonal is subdivided further into $p_{2}\times p_{2}$
sub-blocks, and in the diagonal sub-blocks $Q_{1}$ is replaced by $Q_{2}$.
This procedure of subdivision of the diagonal blocks is repeated, and for $R$
replica symmetry breaking steps, goes down to $p_{R}\times p_{R}$ blocks,
with off-diagonal elements $Q_{R}$. That amounts to take a sequence of $R$
sizes, $p_{0}>p_{1}>p_{2}>\ldots >p_{R}>p_{R+1}$, where by definition $%
p_{0}=n$ and $p_{R+1}=1$, and values $Q_{0}$, $Q_{1}$, $Q_{2}$, $\ldots $, $%
Q_{R}$, having $Q^{aa}=Q_{R+1}=0$. The matrix element%
\begin{equation}
Q^{ab}=Q_{r}  \label{39}
\end{equation}%
depends on the overlap of the replicas $a\cap b=r$, such that, $Q_{r\text{ }%
} $belongs to a block of size $p_{r}$ but not to a sub-block of size $%
p_{r+1} $.

The ansatz can be described equivalently in terms of a tree whose
extremities are the $n$ replicas $a=1,2,\ldots ,n,$ and which foliates at
the various levels $r=0,1,2,\ldots ,R$ with multiplicity $%
n_{r}=p_{r}/p_{r+1} $, as illustrated in figure 1. Each replica is
associated to a string of tree coordinates,%
\begin{equation}
a:\left[ a_{0},a_{1},\ldots ,a_{R}\right]  \label{40}
\end{equation}

\noindent which tells the path to reach replica $a$. Each component takes $%
n_{r}$ values, $a_{r}=1,2,\ldots ,n_{r}$. The overlap of replicas $a$ and $b$
is then defined as%
\begin{eqnarray}
&&\qquad \qquad a\cap b=r,\qquad 0\leq r\leq R+1  \label{41} \\
&&\text{if}\quad a_{0}=b_{0},\,\ldots \,,\,a_{r-1}=b_{r-1},\quad \text{but}%
\quad a_{r}\neq b_{r}  \notag
\end{eqnarray}

\noindent with $a\cap b=R+1$ corresponding to $a=b$. The overlap $a\cap b=r$
represents a kind of hierarchical distance between replicas $a$ and $b$. At
the $r^{th}$ level of hierarchy the order parameter takes the value $%
Q^{ab}=Q_{r}$. The tree displays the geometric properties of the
order-parameter matrix. In particular, it has ultrametricity, that is, given
three replicas $a$, $b$, $c$, the overlaps between these replicas, $a\cap
b=r $, $a\cap c=s$, $b\cap c=t$, either are all equal, or one is larger than
the others, but then these are equal (e.g., $r=s\leq t$).

\bigskip

\begin{center}

\includegraphics[scale=0.8]{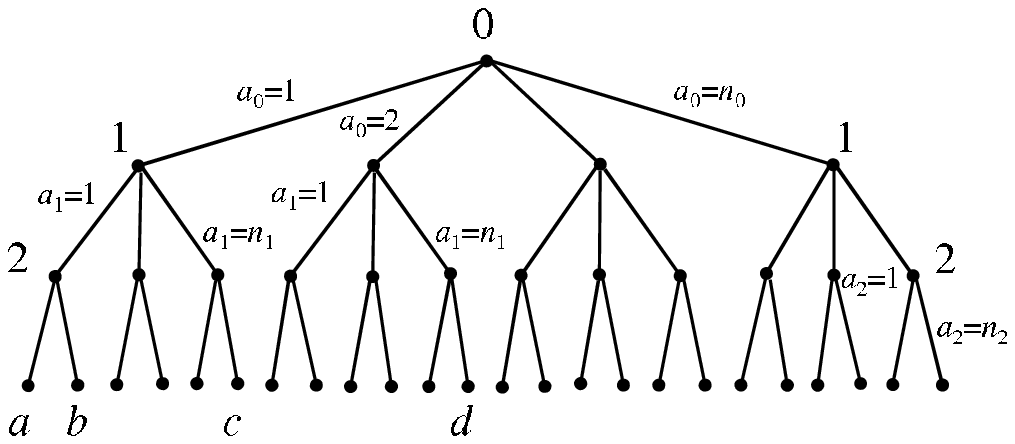}

Figure 1. Tree representation for an $R=2$ RSB ansatz.
\end{center}

\bigskip

Now we consider the Lagrangian term of the fluctuations $\mathcal{L}^{(2)}$,
(9),%
\begin{equation}
\mathcal{L}^{(2)}\mathcal{=}\frac{1}{2}\underset{(ab)(cd)}{\sum }\phi
_{ab}M^{ab,cd}\phi _{cd}  \label{42}
\end{equation}

\noindent again with the dependence on momentum space $\mathbf{p}$ implicit.
The fields are characterized by the overlap of the replicas,%
\begin{equation}
\phi _{ab}=\phi _{r},\quad a\cap b=r  \label{43}
\end{equation}%
with $\phi _{aa}=\phi _{R+1}=0$, and depend on the tree coordinates of the
replicas%
\begin{eqnarray}
\phi _{r} &\equiv &\left[ 
\begin{array}{l}
a_{0}\ldots a_{r-1}a_{r}\ldots a_{R} \\ 
a_{0}\ldots a_{r-1}b_{r}\ldots b_{R}%
\end{array}%
\right] \qquad a_{r}\neq b_{r}  \label{44} \\
&\equiv &\left[ a_{0}\ldots a_{r-1}\left. 
\begin{array}{l}
a_{r}\ldots a_{R} \\ 
b_{r}\ldots b_{R}%
\end{array}%
\right. \right] .  \notag
\end{eqnarray}%
The mass-matrix depends only on the overlaps of the replicas, and can be
parametrized as follows, 
\begin{equation}
M^{ab,cd}=M_{u,v}^{r,s}  \label{45}
\end{equation}

\noindent with%
\begin{eqnarray}
\qquad r &=&a\cap b,\;s=c\cap d  \notag \\
u &=&\max \left( a\cap c,a\cap d\right)  \label{46} \\
v &=&\max \left( b\cap c,b\cap d\right) .  \notag
\end{eqnarray}%
Ultrametricity implies that with four replicas there are generically three
overlaps, i.e., among the overlaps $r$, $s$, $u$, $v$ at least two are
equal; $r$, $s$ are direct-overlaps and $u$, $v$ are cross-overlaps.

The Lagrangian $\mathcal{L}^{(2)}$, (42), is then written as%
\begin{equation}
\mathcal{L}^{(2)}\mathcal{=}\underset{r,s;u,v}{\sum }\underset{\left\{
a,b,c,d\right\} }{\sum }\phi _{r}M_{u,v}^{r,s}\phi _{s}  \label{47}
\end{equation}

\noindent with $0\leq r,s\leq R$, $0\leq u,v\leq R+1$, and where the sum
over the set $\left\{ a,b,c,d\right\} $ depends on the overlaps $r,s,u,v$.
The possible geometries of the tree representation of the mass-matrix, (45),
are presented in figures 2 and 3. We distinguish two sets of contributions:

\noindent $\bullet $ the replicon\textit{\ }(R) configurations, in figure 2,
are characterized by two identical upper indices, $r=s$, and two lower
indices $u$, $v\geq r+1$,%
\begin{equation}
M_{u,v}^{r,s}=M_{u,v}^{r,r};  \label{48}
\end{equation}

\noindent $\bullet $ the longitudinal-anomalous (LA) configurations, in
figure 3, are characterized by a single lower index, $t=\max (u,v)$ (the
other lower index is $r$, $s$, or $t$) and two upper indices $r$, $s$ (where
it may happen, accidently, that $r=s$),%
\begin{equation}
M_{u,v}^{r,s}=M_{t}^{r,s}.  \label{49}
\end{equation}

\noindent The upper indices take values $0,1,\ldots ,R$ and the lower
indices take values $0,1,\ldots ,R+1$. The Lagrangian $\mathcal{L}^{(2)}$
then contains four contributions%
\begin{equation}
\mathcal{L}^{(2)}=\mathcal{L}_{(R)}+\mathcal{L}_{(LA)}^{I}+\mathcal{L}%
_{(LA)}^{II}+\mathcal{L}_{(LA)}^{III}  \label{50}
\end{equation}

\noindent with $\mathcal{L}_{(R)}$ and $\mathcal{L}_{(LA)}^{I}$, $\mathcal{L}%
_{(LA)}^{II}$, $\mathcal{L}_{(LA)}^{III}$ corresponding to the tree
structures in figure 2 and figure 3, respectively.

\bigskip

\begin{center}

\includegraphics[scale=0.7]{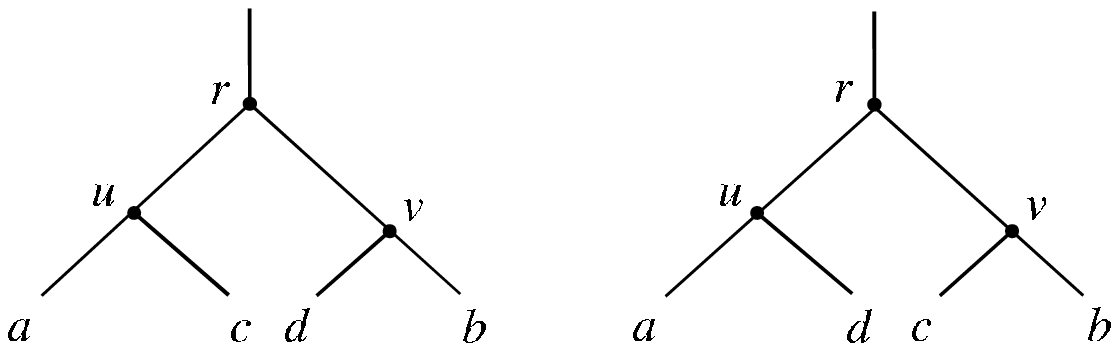}

Figure 2. Tree representation for the replicon sector. The figure shows the
two possible structures compatible with the replicon geometry.

\end{center}

\bigskip

\begin{center}

\includegraphics[scale=0.7]{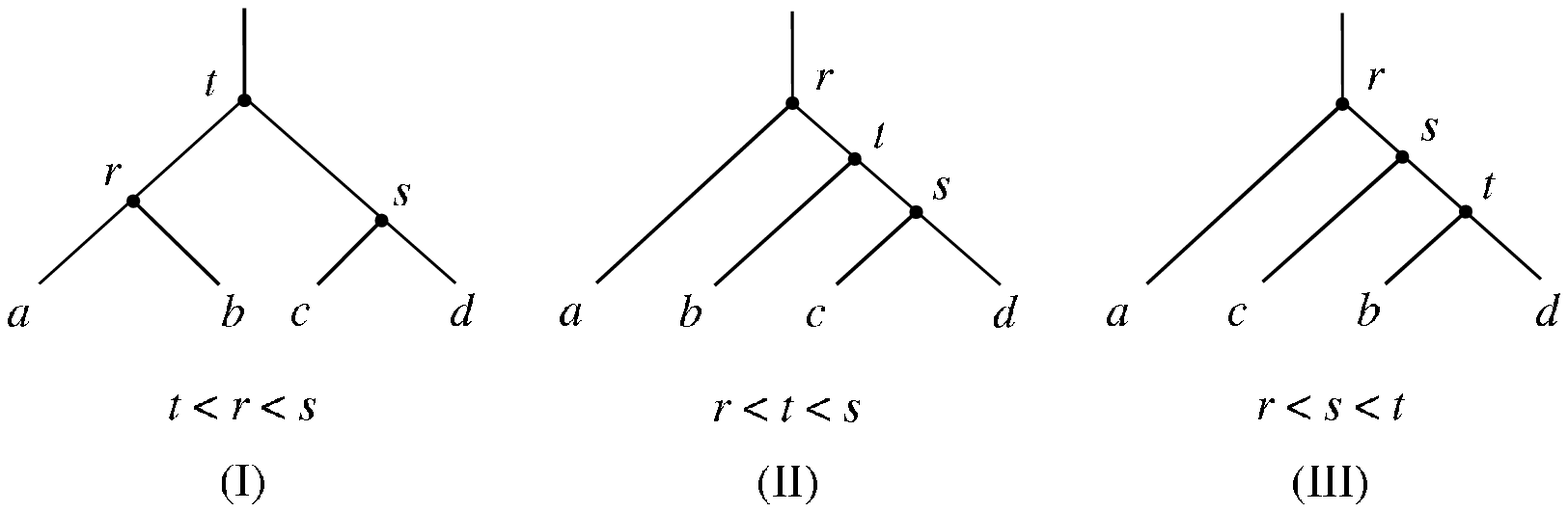}

Figure 3. Tree representation for the longitudinal-anomalous sector.
Exchanging $r$ and $s$ leads to equivalent structures.

\end{center}

\bigskip

We now generalize the RFT introduced in (18). To RFT with respect to replica 
$a$ on a tree, we RFT each of the $\left[ a_{r}\right] $ coordinates of $a$
on the tree. Focusing on $\left[ a_{r}\right] $, one defines%
\begin{eqnarray}
\phi \left[ \hat{a}_{r}\right] &=&\frac{1}{\sqrt{n_{r}}}\sum%
\limits_{a_{r}}e^{-\frac{2\pi i}{n_{r}}a_{r}\hat{a}_{r}}\phi \left[ a_{r}%
\right]  \label{51} \\
\phi \left[ a_{r}\right] &=&\frac{1\,}{\sqrt{n_{r}}}\underset{\hat{a}_{r}}{%
\sum }e^{\frac{2\pi i}{n_{r}}a_{r}\hat{a}_{r}}\phi \left[ \hat{a}_{r}\right]
\notag
\end{eqnarray}

\noindent where $\hat{a}_{r}$ takes $n_{r}=p_{r}/p_{r+1}$ values on the
circle, $\hat{a}_{r}=0,1,2,\ldots ,n_{r}-1,$ $\func{mod}(n_{r})$, having the
relation%
\begin{equation}
\underset{a_{r}}{\sum }e^{\frac{2\pi i}{n_{r}}a_{r}\hat{a}_{r}}=n_{r}\delta
_{\hat{a}_{r},\hat{0}_{r}}.  \label{52}
\end{equation}

\noindent The RFT of the $\left[ a_{r}\right] $ coordinate represents a sum
over the $n_{r}$ values that the component takes at the $r$ level of
foliation, among which there is permutation symmetry. From (51), it follows
that $\phi ^{\ast }\left[ \hat{a}_{r}\right] =\phi \left[ -\hat{a}_{r}\right]
$.

Let us then carry out the steps needed to accomplish the diagonalization of
the Lagrangian $\mathcal{L}^{(2)}$:

\noindent 1. Write the expression for the various contributions to $\mathcal{%
L}^{(2)}$, in (50), which are given by:%
\begin{eqnarray}
\mathcal{L}_{(LA)}^{I} &=&\frac{1}{8}\left\{ \underset{s=r}{\sum^{R}}%
\underset{r=t+1}{\sum^{s}}\underset{t=0}{\sum^{r-1}}M_{t}^{r,s}\times \right.
\label{53} \\
&&\times \sum_{\substack{ \left\{ a,b,c,d\right\}  \\ a_{r}\neq
b_{r},c_{s}\neq d_{s}  \\ a_{t}\neq c_{t}}}%
\begin{array}{l}
\left[ a_{0}\ldots a_{t}\ldots a_{r-1}\left. 
\begin{array}{c}
a_{r}\ldots a_{s}\ldots a_{R} \\ 
b_{r}\ldots b_{s}\ldots b_{R}%
\end{array}%
\right. \right] \\ 
\left[ a_{0}\ldots a_{t-1}c_{t}\ldots c_{r}\ldots c_{s-1}\left. 
\begin{array}{c}
c_{s}\ldots c_{R} \\ 
d_{s}\ldots d_{R}%
\end{array}%
\right. \right]%
\end{array}
\notag
\end{eqnarray}%
\begin{equation*}
\qquad +\underset{s=r+1}{\sum^{R}}\underset{r=0}{\sum^{s-1}}M_{r}^{r,s}\sum 
_{\substack{ \left\{ a,b,c,d\right\}  \\ a_{r}\neq b_{r},c_{s}\neq d_{s}  \\ %
a_{r}\neq c_{r},b_{r}\neq c_{r}}}%
\begin{array}{l}
\left[ a_{0}\ldots a_{r-1}\left. 
\begin{array}{c}
a_{r}\ldots a_{s}\ldots a_{R} \\ 
b_{r}\ldots b_{s}\ldots b_{R}%
\end{array}%
\right. \right] \\ 
\left[ a_{0}\ldots a_{r-1}c_{r}\ldots c_{s-1}\left. 
\begin{array}{c}
c_{s}\ldots c_{R} \\ 
d_{s}\ldots d_{R}%
\end{array}%
\right. \right]%
\end{array}%
\end{equation*}%
$\qquad \qquad $%
\begin{equation*}
+\underset{r=0}{\sum^{R}}M_{r}^{r,r}\sum_{\substack{ \left\{ a,b,c,d\right\} 
\\ a_{r}\neq b_{r},c_{r}\neq d_{r}  \\ a_{r}\neq c_{r},a_{r}\neq d_{r}  \\ %
b_{r}\neq c_{r},b_{r}\neq d_{r}}}%
\begin{array}{l}
\left[ a_{0}\ldots a_{r-1}\left. 
\begin{array}{c}
a_{r}\ldots a_{R} \\ 
b_{r}\ldots b_{R}%
\end{array}%
\right. \right] \\ 
\left[ a_{0}\ldots a_{r-1}\left. 
\begin{array}{c}
c_{r}\ldots c_{R} \\ 
d_{r}\ldots d_{R}%
\end{array}%
\right. \right]%
\end{array}%
\end{equation*}%
\begin{equation*}
\quad +\left. \text{equivalent terms for }t<s<r\text{ and }t=s<r\text{ }%
\right\}
\end{equation*}%
\begin{eqnarray}
\mathcal{L}_{(LA)}^{II} &=&\frac{1}{8}\left\{ \overset{R}{\sum_{s=t+1}}%
\sum_{t=r+1}^{s-1}\sum_{r=0}^{t-1}M_{t}^{r,s}\times \right.  \label{54} \\
&&\times \left( \sum_{\substack{ \left\{ a,b,c,d\right\}  \\ a_{r}\neq
b_{r},c_{s}\neq d_{s}  \\ b_{t}\neq c_{t}}}%
\begin{array}{l}
\left[ a_{0}\ldots a_{r-1}\left. 
\begin{array}{c}
a_{r}\ldots a_{t}\ldots a_{s}\ldots a_{R} \\ 
b_{r}\ldots b_{t}\ldots b_{s}\ldots b_{R}%
\end{array}%
\right. \right] \\ 
\left[ a_{0}\ldots a_{r-1}b_{r}\ldots b_{t-1}c_{t}\ldots c_{s-1}\left. 
\begin{array}{c}
c_{s}\ldots c_{R} \\ 
d_{s}\ldots d_{R}%
\end{array}%
\right. \right]%
\end{array}%
\right.  \notag \\
&&\quad +\left. \sum_{\substack{ \left\{ a,b,c,d\right\}  \\ a_{r}\neq
b_{r},c_{s}\neq d_{s}  \\ a_{t}\neq c_{t}}}%
\begin{array}{l}
\left[ a_{0}\ldots a_{r-1}\left. 
\begin{array}{c}
a_{r}\ldots a_{t}\ldots a_{s}\ldots a_{R} \\ 
b_{r}\ldots b_{t}\ldots b_{s}\ldots b_{R}%
\end{array}%
\right. \right] \\ 
\left[ a_{0}\ldots a_{r}\ldots a_{t-1}c_{t}\ldots c_{s-1}\left. 
\begin{array}{c}
c_{s}\ldots c_{R} \\ 
d_{s}\ldots d_{R}%
\end{array}%
\right. \right]%
\end{array}%
\right)  \notag
\end{eqnarray}%
\begin{eqnarray*}
&&+\overset{R}{\sum_{s=r+1}}\sum_{r=0}^{s-1}M_{s}^{r,s}\left( \sum 
_{\substack{ \left\{ a,b,c,d\right\}  \\ a_{r}\neq b_{r},c_{s}\neq d_{s}  \\ %
b_{s}\neq c_{s},b_{s}\neq d_{s}}}%
\begin{array}{l}
\left[ a_{0}\ldots a_{r-1}\left. 
\begin{array}{c}
a_{r}\ldots a_{s}\ldots a_{R} \\ 
b_{r}\ldots b_{s}\ldots b_{R}%
\end{array}%
\right. \right] \\ 
\left[ a_{0}\ldots a_{r-1}b_{r}\ldots b_{s-1}\left. 
\begin{array}{c}
c_{s}\ldots c_{R} \\ 
d_{s}\ldots d_{R}%
\end{array}%
\right. \right]%
\end{array}%
\right. \qquad \\
&&\qquad \qquad \qquad \quad +\left. \sum_{\substack{ \left\{
a,b,c,d\right\}  \\ a_{r}\neq b_{r},c_{s}\neq d_{s}  \\ a_{s}\neq
c_{s},a_{s}\neq d_{s}}}%
\begin{array}{l}
\left[ a_{0}\ldots a_{r-1}\left. 
\begin{array}{c}
a_{r}\ldots a_{s}\ldots a_{R} \\ 
b_{r}\ldots b_{s}\ldots b_{R}%
\end{array}%
\right. \right] \\ 
\left[ a_{0}\ldots a_{r}\ldots a_{s-1}\left. 
\begin{array}{c}
c_{s}\ldots c_{R} \\ 
d_{s}\ldots d_{R}%
\end{array}%
\right. \right]%
\end{array}%
\right) \qquad
\end{eqnarray*}

\begin{equation*}
\quad +\left. \text{equivalent terms for }s<t<r\text{ and }s<t=r\right\}
\end{equation*}%
\begin{eqnarray}
\mathcal{L}_{(LA)}^{III} &=&\frac{1}{8}\left\{ \underset{t=s+1}{\sum^{R+1}}%
\underset{s=r+1}{\sum^{t-1}}\sum_{r=0}^{s-1}M_{t}^{r,s}\times \right.
\label{55} \\
&&\times \left( \sum_{\substack{ \left\{ a,b,c,d\right\}  \\ a_{r}\neq
b_{r},b_{s}\neq c_{s}  \\ b_{t}\neq d_{t}}}%
\begin{array}{l}
\left[ a_{0}\ldots a_{r-1}\left. 
\begin{array}{c}
a_{r}\ldots a_{s}\ldots a_{t}\ldots a_{R} \\ 
b_{r}\ldots b_{s}\ldots b_{t}\ldots b_{R}%
\end{array}%
\right. \right] \\ 
\left[ a_{0}\ldots a_{r-1}b_{r}\ldots b_{s-1}\left. 
\begin{array}{c}
c_{s}\ldots c_{t-1}c_{t}\ldots c_{R} \\ 
b_{s}\ldots b_{t-1}d_{t}\ldots d_{R}%
\end{array}%
\right. \right]%
\end{array}%
\right.  \notag \\
&&\quad +\sum_{\substack{ \left\{ a,b,c,d\right\}  \\ a_{r}\neq
b_{r},b_{s}\neq d_{s}  \\ b_{t}\neq c_{t}}}%
\begin{array}{l}
\left[ a_{0}\ldots a_{r-1}\left. 
\begin{array}{c}
a_{r}\ldots a_{s}\ldots a_{t}\ldots a_{R} \\ 
b_{r}\ldots b_{s}\ldots b_{t}\ldots b_{R}%
\end{array}%
\right. \right] \\ 
\left[ a_{0}\ldots a_{r-1}b_{r}\ldots b_{s-1}\left. 
\begin{array}{c}
b_{s}\ldots b_{t-1}c_{t}\ldots c_{R} \\ 
d_{s}\ldots d_{t-1}d_{t}\ldots d_{R}%
\end{array}%
\right. \right]%
\end{array}
\notag
\end{eqnarray}%
\begin{eqnarray*}
&&\qquad \qquad +\sum_{\substack{ \left\{ a,b,c,d\right\}  \\ a_{r}\neq
b_{r},a_{s}\neq c_{s}  \\ a_{t}\neq d_{t}}}%
\begin{array}{l}
\left[ a_{0}\ldots a_{r-1}\left. 
\begin{array}{c}
a_{r}\ldots a_{s}\ldots a_{t}\ldots a_{R} \\ 
b_{r}\ldots b_{s}\ldots b_{t}\ldots b_{R}%
\end{array}%
\right. \right] \\ 
\left[ a_{0}\ldots a_{r}\ldots a_{s-1}\left. 
\begin{array}{c}
c_{s}\ldots c_{t-1}c_{t}\ldots c_{R} \\ 
a_{s}\ldots a_{t-1}d_{t}\ldots d_{R}%
\end{array}%
\right. \right]%
\end{array}
\\
&&\qquad \qquad +\left. \sum_{\substack{ \left\{ a,b,c,d\right\}  \\ %
a_{r}\neq b_{r},a_{s}\neq d_{s}  \\ a_{t}\neq c_{t}}}%
\begin{array}{l}
\left[ a_{0}\ldots a_{r-1}\left. 
\begin{array}{c}
a_{r}\ldots a_{s}\ldots a_{t}\ldots a_{R} \\ 
b_{r}\ldots b_{s}\ldots b_{t}\ldots b_{R}%
\end{array}%
\right. \right] \\ 
\left[ a_{0}\ldots a_{r}\ldots a_{s-1}\left. 
\begin{array}{c}
a_{s}\ldots a_{t-1}c_{t}\ldots c_{R} \\ 
d_{s}\ldots d_{t-1}d_{t}\ldots d_{R}%
\end{array}%
\right. \right]%
\end{array}%
\right)
\end{eqnarray*}%
\begin{eqnarray*}
&&+\underset{t=r+1}{\sum^{R+1}}\sum_{r=0}^{t-1}M_{t}^{r,r}\left( \sum 
_{\substack{ \left\{ a,b,c,d\right\}  \\ a_{r}\neq b_{r},b_{r}\neq c_{r}  \\ %
a_{r}\neq c_{r},b_{t}\neq d_{t}}}%
\begin{array}{l}
\left[ a_{0}\ldots a_{r-1}\left. 
\begin{array}{c}
a_{r}\ldots a_{t}\ldots a_{R} \\ 
b_{r}\ldots b_{t}\ldots b_{R}%
\end{array}%
\right. \right] \\ 
\left[ a_{0}\ldots a_{r-1}\left. 
\begin{array}{c}
c_{r}\ldots c_{t-1}c_{t}\ldots c_{R} \\ 
b_{r}\ldots b_{t-1}d_{t}\ldots d_{R}%
\end{array}%
\right. \right]%
\end{array}%
\right. \\
&&\qquad \qquad \qquad \quad +\sum_{\substack{ \left\{ a,b,c,d\right\}  \\ %
a_{r}\neq b_{r},b_{r}\neq d_{r}  \\ a_{r}\neq d_{r},b_{t}\neq c_{t}}}%
\begin{array}{l}
\left[ a_{0}\ldots a_{r-1}\left. 
\begin{array}{c}
a_{r}\ldots a_{t}\ldots a_{R} \\ 
b_{r}\ldots b_{t}\ldots b_{R}%
\end{array}%
\right. \right] \\ 
\left[ a_{0}\ldots a_{r-1}\left. 
\begin{array}{c}
b_{r}\ldots b_{t-1}c_{t}\ldots c_{R} \\ 
d_{r}\ldots d_{t-1}d_{t}\ldots d_{R}%
\end{array}%
\right. \right]%
\end{array}%
\end{eqnarray*}%
\begin{eqnarray*}
&&\qquad \quad \qquad \qquad +\sum_{\substack{ \left\{ a,b,c,d\right\}  \\ %
a_{r}\neq b_{r},a_{r}\neq c_{r}  \\ b_{r}\neq c_{r},a_{t}\neq d_{t}}}%
\begin{array}{l}
\left[ a_{0}\ldots a_{r-1}\left. 
\begin{array}{c}
a_{r}\ldots a_{t}\ldots a_{R} \\ 
b_{r}\ldots b_{t}\ldots b_{R}%
\end{array}%
\right. \right] \\ 
\left[ a_{0}\ldots a_{r-1}\left. 
\begin{array}{c}
c_{r}\ldots c_{t-1}c_{t}\ldots c_{R} \\ 
a_{r}\ldots a_{t-1}d_{t}\ldots d_{R}%
\end{array}%
\right. \right]%
\end{array}
\\
&&\qquad \qquad \qquad \quad +\left. \sum_{\substack{ \left\{
a,b,c,d\right\}  \\ a_{r}\neq b_{r},a_{r}\neq d_{r}  \\ b_{r}\neq
d_{r},a_{t}\neq c_{t}}}%
\begin{array}{l}
\left[ a_{0}\ldots a_{r-1}\left. 
\begin{array}{c}
a_{r}\ldots a_{t}\ldots a_{R} \\ 
b_{r}\ldots b_{t}\ldots b_{R}%
\end{array}%
\right. \right] \\ 
\left[ a_{0}\ldots a_{r-1}\left. 
\begin{array}{c}
a_{r}\ldots a_{t-1}c_{t}\ldots c_{R} \\ 
d_{r}\ldots d_{t-1}d_{t}\ldots d_{R}%
\end{array}%
\right. \right]%
\end{array}%
\right)
\end{eqnarray*}%
\begin{equation*}
+\text{ }\left. \text{equivalent term for }s<r<t\right\}
\end{equation*}%
\begin{eqnarray}
\mathcal{L}_{(R)} &=&\frac{1}{8}\underset{r=0}{\sum^{R}}\underset{u,v=r+1}{%
\sum^{R+1}}M_{u,v}^{r,r}\times  \label{56} \\
&&\times \left( \sum_{\substack{ \left\{ a,b,c,d\right\}  \\ a_{r}\neq b_{r} 
\\ a_{u}\neq c_{u},b_{v}\neq d_{v}}}%
\begin{array}{l}
\left[ a_{0}\ldots a_{r-1}\left. 
\begin{array}{c}
a_{r}\ldots a_{u-1}a_{u}\ldots a_{v-1}a_{v}\ldots a_{R} \\ 
b_{r}\ldots b_{u-1}b_{u}\ldots b_{v-1}b_{v}\ldots b_{R}%
\end{array}%
\right. \right] \\ 
\left[ a_{0}\ldots a_{r-1}\left. 
\begin{array}{c}
a_{r}\ldots a_{u-1}c_{u}\ldots c_{v-1}c_{v}\ldots c_{R} \\ 
b_{r}\ldots b_{u-1}b_{u}\ldots b_{v-1}d_{v}\ldots d_{R}%
\end{array}%
\right. \right]%
\end{array}%
\right.  \notag \\
&&\quad +\left. \sum_{\substack{ \left\{ a,b,c,d\right\}  \\ a_{r}\neq b_{r} 
\\ a_{u}\neq d_{u},b_{v}\neq c_{v}}}%
\begin{array}{l}
\left[ a_{0}\ldots a_{r-1}\left. 
\begin{array}{c}
a_{r}\ldots a_{u-1}a_{u}\ldots a_{v-1}a_{v}\ldots a_{R} \\ 
b_{r}\ldots b_{u-1}b_{u}\ldots b_{v-1}b_{v}\ldots b_{R}%
\end{array}%
\right. \right] \\ 
\left[ a_{0}\ldots a_{r-1}\left. 
\begin{array}{c}
b_{r}\ldots b_{u-1}b_{u}\ldots b_{v-1}c_{v}\ldots c_{R} \\ 
a_{r}\ldots a_{u-1}d_{u}\ldots d_{v-1}d_{v}\ldots d_{R}%
\end{array}%
\right. \right]%
\end{array}%
\right) .  \notag
\end{eqnarray}

\noindent 2. For the cross-overlaps $t$, $u$, $v$, transform the restricted
sums over the tree coordinates into unrestricted sums, which, with
regrouping of terms among the four contributions (53)-(56), leads to:%
\begin{eqnarray}
\mathcal{L}_{(LA)}^{I} &=&\frac{1}{8}\left\{ \underset{s=r}{\sum^{R}}%
\underset{r=t}{\sum^{s}}\underset{t=0}{\sum^{r}}\right.
(M_{t}^{r,s}-M_{t-1}^{r,s})\times  \label{57} \\
&&\times \sum_{\substack{ \left\{ a,b,c,d\right\}  \\ a_{r}\neq
b_{r},c_{s}\neq d_{s}}}%
\begin{array}{l}
\left[ a_{0}\ldots a_{t-1}a_{t}\ldots a_{r-1}\left. 
\begin{array}{c}
a_{r}\ldots a_{s}\ldots a_{R} \\ 
b_{r}\ldots b_{s}\ldots b_{R}%
\end{array}%
\right. \right] \\ 
\left[ a_{0}\ldots a_{t-1}c_{t}\ldots c_{r}\ldots c_{s-1}\left. 
\begin{array}{c}
c_{s}\ldots c_{R} \\ 
d_{s}\ldots d_{R}%
\end{array}%
\right. \right]%
\end{array}
\notag
\end{eqnarray}

\begin{equation*}
+\left. \text{ equivalent term for }t\leq s<r\right\}
\end{equation*}

\noindent fixing $M_{-1}^{r,s}=0$,%
\begin{eqnarray}
\mathcal{L}_{(LA)}^{II} &=&\frac{1}{8}\left\{ \overset{R}{\sum_{s=t}}%
\sum_{t=r+1}^{s}\sum_{r=0}^{t-1}\right. \left(
M_{t}^{r,s}-M_{t-1}^{r,s}\right) \times  \label{58} \\
&&\times \sum_{\substack{ \left\{ a,b,c,d\right\}  \\ a_{r}\neq
b_{r},c_{s}\neq d_{s}}}\left( 
\begin{array}{l}
\left[ a_{0}\ldots a_{r-1}\left. 
\begin{array}{c}
a_{r}\ldots a_{t}\ldots a_{s}\ldots a_{R} \\ 
b_{r}\ldots b_{t}\ldots b_{s}\ldots b_{R}%
\end{array}%
\right. \right] \\ 
\left[ a_{0}\ldots a_{r-1}b_{r}\ldots b_{t-1}c_{t}\ldots c_{s-1}\left. 
\begin{array}{c}
c_{s}\ldots c_{R} \\ 
d_{s}\ldots d_{R}%
\end{array}%
\right. \right]%
\end{array}%
\right.  \notag \\
&&\qquad \qquad \qquad +\left. 
\begin{array}{l}
\left[ a_{0}\ldots a_{r-1}\left. 
\begin{array}{c}
a_{r}\ldots a_{t}\ldots a_{s}\ldots a_{R} \\ 
b_{r}\ldots b_{t}\ldots b_{s}\ldots b_{R}%
\end{array}%
\right. \right] \\ 
\left[ a_{0}\ldots a_{r}\ldots a_{t-1}c_{t}\ldots c_{s-1}\left. 
\begin{array}{c}
c_{s}\ldots c_{R} \\ 
d_{s}\ldots d_{R}%
\end{array}%
\right. \right]%
\end{array}%
\right)  \notag
\end{eqnarray}%
\begin{equation*}
+\left. \text{equivalent term for }s<t\leq r\right\}
\end{equation*}%
\begin{eqnarray}
\mathcal{L}_{(LA)}^{III} &=&\frac{1}{8}\left\{ \underset{t=s+1}{\sum^{R+1}}%
\underset{s=r}{\sum^{t-1}}\sum_{r=0}^{s}\right. \left(
M_{t}^{r,s}-M_{t-1}^{r,s}\right) \times  \label{59} \\
&&\times \left( \sum_{\substack{ \left\{ a,b,c,d\right\}  \\ a_{r}\neq
b_{r},b_{s}\neq c_{s}}}%
\begin{array}{l}
\left[ a_{0}\ldots a_{r-1}\left. 
\begin{array}{c}
a_{r}\ldots a_{s}\ldots a_{t}\ldots a_{R} \\ 
b_{r}\ldots b_{s}\ldots b_{t}\ldots b_{R}%
\end{array}%
\right. \right] \\ 
\left[ a_{0}\ldots a_{r-1}b_{r}\ldots b_{s-1}\left. 
\begin{array}{c}
c_{s}\ldots c_{t-1}c_{t}\ldots c_{R} \\ 
b_{s}\ldots b_{t-1}d_{t}\ldots d_{R}%
\end{array}%
\right. \right]%
\end{array}%
\right.  \notag \\
&&+\sum_{\substack{ \left\{ a,b,c,d\right\}  \\ a_{r}\neq b_{r},b_{s}\neq
d_{s}}}%
\begin{array}{l}
\left[ a_{0}\ldots a_{r-1}\left. 
\begin{array}{c}
a_{r}\ldots a_{s}\ldots a_{t}\ldots a_{R} \\ 
b_{r}\ldots b_{s}\ldots b_{t}\ldots b_{R}%
\end{array}%
\right. \right] \\ 
\left[ a_{0}\ldots a_{r-1}b_{r}\ldots b_{s-1}\left. 
\begin{array}{c}
b_{s}\ldots b_{t-1}c_{t}\ldots c_{R} \\ 
d_{s}\ldots d_{t-1}d_{t}\ldots d_{R}%
\end{array}%
\right. \right]%
\end{array}
\notag
\end{eqnarray}%
\begin{eqnarray*}
&&\qquad \qquad \quad +\sum_{\substack{ \left\{ a,b,c,d\right\}  \\ %
a_{r}\neq b_{r},a_{s}\neq c_{s}}}%
\begin{array}{l}
\left[ a_{0}\ldots a_{r-1}\left. 
\begin{array}{c}
a_{r}\ldots a_{s}\ldots a_{t}\ldots a_{R} \\ 
b_{r}\ldots b_{s}\ldots b_{t}\ldots b_{R}%
\end{array}%
\right. \right] \\ 
\left[ a_{0}\ldots a_{r-1}a_{r}\ldots a_{s-1}\left. 
\begin{array}{c}
c_{s}\ldots c_{t-1}c_{t}\ldots c_{R} \\ 
a_{s}\ldots a_{t-1}d_{t}\ldots d_{R}%
\end{array}%
\right. \right]%
\end{array}
\\
&&\qquad \qquad \quad +\left. \sum_{\substack{ \left\{ a,b,c,d\right\}  \\ %
a_{r}\neq b_{r},a_{s}\neq d_{s}}}%
\begin{array}{l}
\left[ a_{0}\ldots a_{r-1}\left. 
\begin{array}{c}
a_{r}\ldots a_{s}\ldots a_{t}\ldots a_{R} \\ 
b_{r}\ldots b_{s}\ldots b_{t}\ldots b_{R}%
\end{array}%
\right. \right] \\ 
\left[ a_{0}\ldots a_{r-1}a_{r}\ldots a_{s-1}\left. 
\begin{array}{c}
a_{s}\ldots a_{t-1}c_{t}\ldots c_{R} \\ 
d_{s}\ldots d_{t-1}d_{t}\ldots d_{R}%
\end{array}%
\right. \right]%
\end{array}%
\right)
\end{eqnarray*}%
\begin{equation*}
+\text{ }\left. \text{equivalent term for }s<t<r\right\}
\end{equation*}%
\begin{eqnarray}
\mathcal{L}_{(R)} &=&\frac{1}{8}\underset{r=0}{\sum^{R}}\underset{u,v=r+1}{%
\sum^{R+1}}\left(
M_{u,v}^{r,r}-M_{u-1,v}^{r,r}-M_{u,v-1}^{r,r}+M_{u-1,v-1}^{r,r}\right) \times
\label{60} \\
&&\times \sum_{\substack{ \left\{ a,b,c,d\right\}  \\ a_{r}\neq b_{r}}}%
\left( 
\begin{array}{l}
\left[ a_{0}\ldots a_{r-1}\left. 
\begin{array}{c}
a_{r}\ldots a_{u-1}a_{u}\ldots a_{v-1}a_{v}\ldots a_{R} \\ 
b_{r}\ldots b_{u-1}b_{u}\ldots b_{v-1}b_{v}\ldots b_{R}%
\end{array}%
\right. \right] \\ 
\left[ a_{0}\ldots a_{r-1}\left. 
\begin{array}{c}
a_{r}\ldots a_{u-1}c_{u}\ldots c_{v-1}c_{v}\ldots c_{R} \\ 
b_{r}\ldots b_{u-1}b_{u}\ldots b_{v-1}d_{v}\ldots d_{R}%
\end{array}%
\right. \right]%
\end{array}%
\right.  \notag \\
&&\qquad \qquad +\left. 
\begin{array}{l}
\left[ a_{0}\ldots a_{r-1}\left. 
\begin{array}{c}
a_{r}\ldots a_{u-1}a_{u}\ldots a_{v-1}a_{v}\ldots a_{R} \\ 
b_{r}\ldots b_{u-1}b_{u}\ldots b_{v-1}b_{v}\ldots b_{R}%
\end{array}%
\right. \right] \\ 
\left[ a_{0}\ldots a_{r-1}\left. 
\begin{array}{c}
b_{r}\ldots b_{u-1}b_{u}\ldots b_{v-1}c_{v}\ldots c_{R} \\ 
a_{r}\ldots a_{u-1}d_{u}\ldots d_{v-1}d_{v}\ldots d_{R}%
\end{array}%
\right. \right]%
\end{array}%
\right) .  \notag
\end{eqnarray}

\noindent 3. Perform the RFT\ on all the tree coordinates $a$, $b$, $c$, $d$
of the replicas, which leads to:%
\begin{eqnarray}
\mathcal{L}_{(LA)}^{I} &=&\frac{1}{8}\left\{ \underset{s=r}{\sum^{R}}%
\underset{r=t}{\sum^{s}}\underset{t=0}{\sum^{r}}\right. \sum_{\left\{ \hat{%
\gamma}\right\} }\sqrt{\delta _{r}}\sqrt{\delta _{s}}%
p_{t}(M_{t}^{r,s}-M_{t-1}^{r,s})\times  \label{61} \\
&&\times 
\begin{array}{l}
\left[ \hat{\gamma}_{0}\ldots \hat{\gamma}_{t-1}\hat{0}_{t}\ldots \hat{0}%
_{r-1}\left( 
\begin{array}{c}
\hat{0}_{r} \\ 
\hat{0}_{r}%
\end{array}%
\right) \left. 
\begin{array}{c}
\hat{0}_{r+1}\ldots \hat{0}_{s}\ldots \hat{0}_{R} \\ 
\hat{0}_{r+1}\ldots \hat{0}_{s}\ldots \hat{0}_{R}%
\end{array}%
\right. \right] _{N} \\ 
\left[ \hat{\gamma}_{0}\ldots \hat{\gamma}_{t-1}\hat{0}_{t}\ldots \hat{0}%
_{r}\ldots \hat{0}_{s-1}\left( 
\begin{array}{c}
\hat{0}_{s} \\ 
\hat{0}_{s}%
\end{array}%
\right) \left. 
\begin{array}{c}
\hat{0}_{s+1}\ldots \hat{0}_{R} \\ 
\hat{0}_{s+1}\ldots \hat{0}_{R}%
\end{array}%
\right. \right] _{N}^{\ast }%
\end{array}
\notag
\end{eqnarray}%
\begin{equation*}
+\left. \text{ equivalent term for }t\leq s<r\right\}
\end{equation*}%
\begin{eqnarray}
\mathcal{L}_{(LA)}^{II} &=&\frac{1}{8}\left\{ \overset{R}{\sum_{s=t}}%
\sum_{t=r+1}^{s}\sum_{r=0}^{t-1}\right. \sum_{\left\{ \hat{\gamma}\right\} }%
\sqrt{\delta _{r}}\sqrt{\delta _{s}}p_{t}\left(
M_{t}^{r,s}-M_{t-1}^{r,s}\right) \times  \label{62} \\
&&\times \left( \left[ \hat{\gamma}_{0}\ldots \hat{\gamma}_{r-1}\left( 
\begin{array}{c}
\hat{0}_{r} \\ 
\hat{\gamma}_{r}%
\end{array}%
\right) \left. 
\begin{array}{c}
\hat{0}_{r+1}\ldots \hat{0}_{t-1}\hat{0}_{t}\ldots \hat{0}_{s}\ldots \hat{0}%
_{R} \\ 
\hat{\gamma}_{r+1}\ldots \hat{\gamma}_{t-1}\hat{0}_{t}\ldots \hat{0}%
_{s}\ldots \hat{0}_{R}%
\end{array}%
\right. \right] _{N}\right.  \notag \\
&&\qquad +\left. \left[ \hat{\gamma}_{0}\ldots \hat{\gamma}_{r-1}\left( 
\begin{array}{c}
\hat{\gamma}_{r} \\ 
\hat{0}_{r}%
\end{array}%
\right) \left. 
\begin{array}{c}
\hat{\gamma}_{r+1}\ldots \hat{\gamma}_{t-1}\hat{0}_{t}\ldots \hat{0}%
_{s}\ldots \hat{0}_{R} \\ 
\hat{0}_{r+1}\ldots \hat{0}_{t-1}\hat{0}_{t}\ldots \hat{0}_{s}\ldots \hat{0}%
_{R}%
\end{array}%
\right. \right] _{N}\right)  \notag \\
&&\times \left[ \hat{\gamma}_{0}\ldots \hat{\gamma}_{r}\ldots \hat{\gamma}%
_{t-1}\hat{0}_{t}\ldots \hat{0}_{s-1}\left( 
\begin{array}{c}
\hat{0}_{s} \\ 
\hat{0}_{s}%
\end{array}%
\right) \left. 
\begin{array}{c}
\hat{0}_{s+1}\ldots \hat{0}_{R} \\ 
\hat{0}_{s+1}\ldots \hat{0}_{R}%
\end{array}%
\right. \right] _{N}^{\ast }  \notag
\end{eqnarray}%
\begin{equation*}
+\text{ }\left. \text{equivalent term for }s<t\leq r\right\}
\end{equation*}%
\begin{eqnarray}
\mathcal{L}_{(LA)}^{III} &=&\frac{1}{8}\left\{ \underset{t=s+1}{\sum^{R+1}}%
\underset{s=r}{\sum^{t-1}}\sum_{r=0}^{s}\right. \sum_{\left\{ \hat{\gamma}%
\right\} }\sqrt{\delta _{r}}\sqrt{\delta _{s}}p_{t}\left(
M_{t}^{r,s}-M_{t-1}^{r,s}\right) \times  \label{63} \\
&&\times \left( \left[ \hat{\gamma}_{0}\ldots \hat{\gamma}_{r-1}\left( 
\begin{array}{c}
\hat{0}_{r} \\ 
\hat{\gamma}_{r}%
\end{array}%
\right) \left. 
\begin{array}{c}
\hat{0}_{r+1}\ldots \hat{0}_{s}\ldots \hat{0}_{t-1}\hat{0}_{t}\ldots \hat{0}%
_{R} \\ 
\hat{\gamma}_{r+1}\ldots \hat{\gamma}_{s}\ldots \hat{\gamma}_{t-1}\hat{0}%
_{t}\ldots \hat{0}_{R}%
\end{array}%
\right. \right] _{N}\right.  \notag \\
&&\qquad +\left. \left[ \hat{\gamma}_{0}\ldots \hat{\gamma}_{r-1}\left( 
\begin{array}{c}
\hat{\gamma}_{r} \\ 
\hat{0}_{r}%
\end{array}%
\right) \left. 
\begin{array}{c}
\hat{\gamma}_{r+1}\ldots \hat{\gamma}_{s}\ldots \hat{\gamma}_{t-1}\hat{0}%
_{t}\ldots \hat{0}_{R} \\ 
\hat{0}_{r+1}\ldots \hat{0}_{s}\ldots \hat{0}_{t-1}\hat{0}_{t}\ldots \hat{0}%
_{R}%
\end{array}%
\right. \right] _{N}\right)  \notag \\
&&\times \left( \left[ \hat{\gamma}_{0}\ldots \hat{\gamma}_{r}\ldots \hat{%
\gamma}_{s-1}\left( 
\begin{array}{c}
\hat{0}_{s} \\ 
\hat{\gamma}_{s}%
\end{array}%
\right) \left. 
\begin{array}{c}
\hat{0}_{s+1}\ldots \hat{0}_{t-1}\hat{0}_{t}\ldots \hat{0}_{R} \\ 
\hat{\gamma}_{s+1}\ldots \hat{\gamma}_{t-1}\hat{0}_{t}\ldots \hat{0}_{R}%
\end{array}%
\right. \right] _{N}^{\ast }\right.  \notag \\
&&\qquad +\left. \left[ \hat{\gamma}_{0}\ldots \hat{\gamma}_{r}\ldots \hat{%
\gamma}_{s-1}\left( 
\begin{array}{c}
\hat{\gamma}_{s} \\ 
\hat{0}_{s}%
\end{array}%
\right) \left. 
\begin{array}{c}
\hat{\gamma}_{s+1}\ldots \hat{\gamma}_{t-1}\hat{0}_{t}\ldots \hat{0}_{R} \\ 
\hat{0}_{s+1}\ldots \hat{0}_{t-1}\hat{0}_{t}\ldots \hat{0}_{R}%
\end{array}%
\right. \right] _{N}^{\ast }\right)  \notag
\end{eqnarray}%
\begin{equation*}
+\text{ }\left. \text{equivalent term for }s<t<r\right\}
\end{equation*}%
\begin{eqnarray}
&&\mathcal{L}_{(R)}=\frac{1}{8}\underset{r=0}{\sum^{R}}\underset{u,v=r+1}{%
\sum^{R+1}}\sum_{\left\{ \hat{\gamma},\hat{\mu},\hat{\nu}\right\} }\left( 
\frac{n_{r}-1}{n_{r}}\right) \times  \label{64} \\
&&\qquad \quad \times p_{u}p_{v}\left(
M_{u,v}^{r,r}-M_{u-1,v}^{r,r}-M_{u,v-1}^{r,r}+M_{u-1,v-1}^{r,r}\right) \times
\notag \\
&&\times \left( \left[ \hat{\gamma}_{0}\ldots \hat{\gamma}_{r-1}\left( 
\begin{array}{c}
\hat{\mu}_{r} \\ 
\hat{\gamma}_{r}-\hat{\mu}_{r}%
\end{array}%
\right) \left. 
\begin{array}{c}
\hat{\mu}_{r+1}\ldots \hat{\mu}_{u-1}\hat{0}_{u}\ldots \hat{0}_{v-1}\hat{0}%
_{v}\ldots \hat{0}_{R} \\ 
\hat{\nu}_{r+1}\ldots \hat{\nu}_{u-1}\hat{\nu}_{u}\ldots \hat{\nu}_{v-1}\hat{%
0}_{v}\ldots \hat{0}_{R}%
\end{array}%
\right. \right] _{N}\right)  \notag \\
&&\times \left( \left[ \hat{\gamma}_{0}\ldots \hat{\gamma}_{r-1}\left( 
\begin{array}{c}
\hat{\mu}_{r} \\ 
\hat{\gamma}_{r}-\hat{\mu}_{r}%
\end{array}%
\right) \left. 
\begin{array}{c}
\hat{\mu}_{r+1}\ldots \hat{\mu}_{u-1}\hat{0}_{u}\ldots \hat{0}_{v-1}\hat{0}%
_{v}\ldots \hat{0}_{R} \\ 
\hat{\nu}_{r+1}\ldots \hat{\nu}_{u-1}\hat{\nu}_{u}\ldots \hat{\nu}_{v-1}\hat{%
0}_{v}\ldots \hat{0}_{R}%
\end{array}%
\right. \right] _{N}^{\ast }\right.  \notag \\
&&+\left. \left[ \hat{\gamma}_{0}\ldots \hat{\gamma}_{r-1}\left( 
\begin{array}{c}
\hat{\gamma}_{r}-\hat{\mu}_{r} \\ 
\hat{\mu}_{r}%
\end{array}%
\right) \left. 
\begin{array}{c}
\hat{\nu}_{r+1}\ldots \hat{\nu}_{u-1}\hat{\nu}_{u}\ldots \hat{\nu}_{v-1}\hat{%
0}_{v}\ldots \hat{0}_{R} \\ 
\hat{\mu}_{r+1}\ldots \hat{\mu}_{u-1}\hat{0}_{u}\ldots \hat{0}_{v-1}\hat{0}%
_{v}\ldots \hat{0}_{R}%
\end{array}%
\right. \right] _{N}^{\ast }\right)  \notag
\end{eqnarray}

\noindent \noindent where we define%
\begin{equation}
\delta _{r}=\left( p_{r}-p_{r+1}\right)  \label{65}
\end{equation}

\noindent and use the notation%
\begin{equation}
\left[ \hat{\gamma}_{r}\right] =\underset{\hat{a}_{r}}{\sum }\left[ 
\begin{array}{c}
\hat{a}_{r} \\ 
\hat{\gamma}_{r}-\hat{a}_{r}%
\end{array}%
\right] ,  \label{66}
\end{equation}

\noindent normalized, $\left[ \hat{\gamma}_{r}\right] _{N}=\frac{1}{\sqrt{%
n_{r}}}\left[ \hat{\gamma}_{r}\right] $. The restrictions on the tree
coordinates associate with the direct-overlaps $r$, $s$ are incorporated in
(61)-(64), by introducing the \textit{marker }definition%
\begin{equation}
\left( 
\begin{array}{c}
\hat{\mu}_{r} \\ 
\hat{\gamma}_{r}-\hat{\mu}_{r}%
\end{array}%
\right) =\left[ 
\begin{array}{c}
\hat{\mu}_{r} \\ 
\hat{\gamma}_{r}-\hat{\mu}_{r}%
\end{array}%
\right] -\frac{1}{n_{r}}\left[ \hat{\gamma}_{r}\right]  \label{67}
\end{equation}

\noindent which corresponds to the RFT\ of $\left[ 
\begin{array}{c}
a_{r} \\ 
b_{r}%
\end{array}%
\right] _{a_{r}\neq b_{r}}$. The \textit{marker} has the property%
\begin{equation}
\underset{\hat{\mu}_{r}}{\sum }\left( 
\begin{array}{c}
\hat{\mu}_{r} \\ 
\hat{\gamma}_{r}-\hat{\mu}_{r}%
\end{array}%
\right) =0,  \label{68}
\end{equation}

\noindent and normalization,%
\begin{equation}
\left( 
\begin{array}{c}
\hat{\mu}_{r} \\ 
\hat{\gamma}_{r}-\hat{\mu}_{r}%
\end{array}%
\right) =\sqrt{\frac{n_{r}-1}{n_{r}}}\left( 
\begin{array}{c}
\hat{\mu}_{r} \\ 
\hat{\gamma}_{r}-\hat{\mu}_{r}%
\end{array}%
\right) _{N}.  \label{69}
\end{equation}%
We have also used the relation%
\begin{eqnarray}
\underset{\hat{\gamma}_{r},\hat{\mu}_{r}}{\sum }\left( 
\begin{array}{c}
\hat{\mu}_{r} \\ 
\hat{\gamma}_{r}-\hat{\mu}_{r}%
\end{array}%
\right) \left( 
\begin{array}{c}
\hat{\mu}_{r} \\ 
\hat{\gamma}_{r}-\hat{\mu}_{r}%
\end{array}%
\right) ^{\ast } &=&\underset{\hat{\gamma}_{r},\hat{\mu}_{r}}{\sum }\left[ 
\begin{array}{c}
\hat{\mu}_{r} \\ 
\hat{\gamma}_{r}-\hat{\mu}_{r}%
\end{array}%
\right] \left[ 
\begin{array}{c}
\hat{\mu}_{r} \\ 
\hat{\gamma}_{r}-\hat{\mu}_{r}%
\end{array}%
\right] ^{\ast }  \notag \\
&&-\underset{\hat{\gamma}_{r}}{\sum }\left[ \hat{\gamma}_{r}\right] _{N}%
\left[ \hat{\gamma}_{r}\right] _{N}^{\ast }.  \label{70}
\end{eqnarray}

\noindent 4. Separate in the sums over the tree coordinates of the replicas,
the $\hat{0}$ components from the $\hat{\gamma}^{\prime }\neq \hat{0}$, $%
\hat{\mu}^{\prime }\neq \hat{0}$, $\hat{\nu}^{\prime }\neq \hat{0}$
components.

For $\mathcal{L}_{(LA)}^{I}$, $\mathcal{L}_{(LA)}^{II}$, $\mathcal{L}%
_{(LA)}^{III}$ one obtains:%
\begin{eqnarray}
\mathcal{L}_{(LA)}^{I} &=&\frac{1}{16}\left\{ \underset{s=r}{\sum^{R}}%
\underset{r=t}{\sum^{s}}\underset{t=0}{\sum^{r}}\right. \sum_{\left\{ \hat{%
\gamma}\right\} }\sqrt{\delta _{r}^{(t-1)}}\hat{M}_{t}^{r,s}\sqrt{\delta
_{s}^{(t-1)}}\times  \label{71} \\
&&\times 
\begin{array}{l}
\left[ \hat{\gamma}_{0}\ldots \hat{\gamma}_{t-1}^{\prime }\hat{0}_{t}\ldots 
\hat{0}_{r-1}\left( 
\begin{array}{c}
\hat{0}_{r} \\ 
\hat{0}_{r}%
\end{array}%
\right) \left. 
\begin{array}{c}
\hat{0}_{r+1}\ldots \hat{0}_{s}\ldots \hat{0}_{R} \\ 
\hat{0}_{r+1}\ldots \hat{0}_{s}\ldots \hat{0}_{R}%
\end{array}%
\right. \right] _{SN} \\ 
\left[ \hat{\gamma}_{0}\ldots \hat{\gamma}_{t-1}^{\prime }\hat{0}_{t}\ldots 
\hat{0}_{r}\ldots \hat{0}_{s-1}\left( 
\begin{array}{c}
\hat{0}_{s} \\ 
\hat{0}_{s}%
\end{array}%
\right) \left. 
\begin{array}{c}
\hat{0}_{s+1}\ldots \hat{0}_{R} \\ 
\hat{0}_{s+1}\ldots \hat{0}_{R}%
\end{array}%
\right. \right] _{SN}^{\ast }%
\end{array}
\notag
\end{eqnarray}%
\begin{equation*}
+\left. \text{ equivalent term for }t\leq s<r\right\}
\end{equation*}%
\begin{eqnarray}
\mathcal{L}_{(LA)}^{II} &=&\frac{1}{16}\left\{ \overset{R}{\sum_{s=t}}%
\underset{t=r+1}{\sum^{s}}\sum_{r=0}^{t-1}\right. \sum_{\left\{ \hat{\gamma}%
\right\} }\sqrt{\delta _{r}^{(t-1)}}\hat{M}_{t}^{r,s}\sqrt{\delta
_{s}^{(t-1)}}\times  \label{72} \\
&&\times 
\begin{array}{l}
\left[ \hat{\gamma}_{0}\ldots \hat{\gamma}_{r-1}\left( 
\begin{array}{c}
\hat{0}_{r} \\ 
\hat{\gamma}_{r}%
\end{array}%
\right) \left. 
\begin{array}{c}
\hat{0}_{r+1}\ldots \hat{0}_{t-1}\hat{0}_{t}\ldots \hat{0}_{s}\ldots \hat{0}%
_{R} \\ 
\hat{\gamma}_{r+1}\ldots \hat{\gamma}_{t-1}^{\prime }\hat{0}_{t}\ldots \hat{0%
}_{s}\ldots \hat{0}_{R}%
\end{array}%
\right. \right] _{SN} \\ 
\left[ \hat{\gamma}_{0}\ldots \hat{\gamma}_{r}\ldots \hat{\gamma}%
_{t-1}^{\prime }\hat{0}_{t}\ldots \hat{0}_{s-1}\left( 
\begin{array}{c}
\hat{0}_{s} \\ 
\hat{0}_{s}%
\end{array}%
\right) \left. 
\begin{array}{c}
\hat{0}_{s+1}\ldots \hat{0}_{R} \\ 
\hat{0}_{s+1}\ldots \hat{0}_{R}%
\end{array}%
\right. \right] _{SN}^{\ast }%
\end{array}
\notag
\end{eqnarray}%
\begin{equation*}
+\text{ }\left. \text{equivalent term for }s<t\leq r\right\}
\end{equation*}%
\begin{eqnarray}
\mathcal{L}_{(LA)}^{III} &=&\frac{1}{16}\left\{ \underset{t=s+1}{\sum^{R+1}}%
\underset{s=r}{\sum^{t-1}}\sum_{r=0}^{s}\right. \sum_{\left\{ \hat{\gamma}%
\right\} }\sqrt{\delta _{r}^{(t-1)}}\hat{M}_{t}^{r,s}\sqrt{\delta
_{s}^{(t-1)}}\times  \label{73} \\
&&\times 
\begin{array}{l}
\left[ \hat{\gamma}_{0}\ldots \hat{\gamma}_{r-1}\left( 
\begin{array}{c}
\hat{0}_{r} \\ 
\hat{\gamma}_{r}%
\end{array}%
\right) \left. 
\begin{array}{c}
\hat{0}_{r+1}\ldots \hat{0}_{s}\ldots \hat{0}_{t-1}\hat{0}_{t}\ldots \hat{0}%
_{R} \\ 
\hat{\gamma}_{r+1}\ldots \hat{\gamma}_{s}\ldots \hat{\gamma}_{t-1}^{\prime }%
\hat{0}_{t}\ldots \hat{0}_{R}%
\end{array}%
\right. \right] _{SN} \\ 
\left[ \hat{\gamma}_{0}\ldots \hat{\gamma}_{r}\ldots \hat{\gamma}%
_{s-1}\left( 
\begin{array}{c}
\hat{0}_{s} \\ 
\hat{\gamma}_{s}%
\end{array}%
\right) \left. 
\begin{array}{c}
\hat{0}_{s+1}\ldots \hat{0}_{t-1}\hat{0}_{t}\ldots \hat{0}_{R} \\ 
\hat{\gamma}_{s+1}\ldots \hat{\gamma}_{t-1}^{\prime }\hat{0}_{t}\ldots \hat{0%
}_{R}%
\end{array}%
\right. \right] _{SN}^{\ast }%
\end{array}
\notag
\end{eqnarray}%
\begin{equation*}
+\text{ }\left. \text{equivalent term for }s<t<r\right\}
\end{equation*}

\noindent having, at the end, symmetrized the fields at the \textit{marker},%
\begin{equation}
\left( 
\begin{array}{c}
\hat{0}_{r} \\ 
\hat{\gamma}_{r}%
\end{array}%
\right) _{S}=\frac{1}{2}\left[ \left( 
\begin{array}{c}
\hat{0}_{r} \\ 
\hat{\gamma}_{r}%
\end{array}%
\right) _{N}+\left( 
\begin{array}{c}
\hat{\gamma}_{r} \\ 
\hat{0}_{r}%
\end{array}%
\right) _{N}\right] ,  \label{74}
\end{equation}

\noindent and normalized, for $t=r+1$,%
\begin{equation}
\left( 
\begin{array}{c}
\hat{0}_{r} \\ 
\hat{\gamma}_{r}%
\end{array}%
\right) _{S}=\sqrt{\frac{\delta _{r}^{(t-1)}}{2\delta _{r}}}\left( 
\begin{array}{c}
\hat{0}_{r} \\ 
\hat{\gamma}_{r}%
\end{array}%
\right) _{SN}  \label{75}
\end{equation}%
with%
\begin{equation}
\delta _{r}^{(l)}=p_{r}^{(l)}-p_{r+1}^{(l)}  \label{76}
\end{equation}%
\begin{equation}
p_{r}^{(l)}=\left\{ 
\begin{array}{c}
p_{r}\,,\quad r\leq l \\ 
2p_{r}\,,\quad r>l%
\end{array}%
\right.  \label{77}
\end{equation}

\noindent and, for $t>r+1$,%
\begin{equation}
\left( 
\begin{array}{c}
\hat{0}_{r} \\ 
\hat{\gamma}_{r}%
\end{array}%
\right) _{S}=\frac{1}{\sqrt{2}}\left( 
\begin{array}{c}
\hat{0}_{r} \\ 
\hat{\gamma}_{r}%
\end{array}%
\right) _{SN}.  \label{78}
\end{equation}%
In (71)-(73) $\hat{M}_{t}^{r,s}$ is the mass RFT [21], 
\begin{equation}
\hat{M}_{t}^{r,s}=\underset{k=t}{\sum^{R+1}}p_{k}^{(r,s)}\left(
M_{k}^{r,s}-M_{k-1}^{r,s}\right)  \label{79}
\end{equation}

\noindent with%
\begin{eqnarray}
p_{k}^{(r,s)} &=&p_{k},\quad k\leq r\leq s  \notag \\
p_{k}^{(r,s)} &=&2p_{k},\quad r<k\leq s  \label{80} \\
p_{k}^{(r,s)} &=&4p_{k},\quad r\leq s<k  \notag
\end{eqnarray}

\noindent the inverse\ transform being given by%
\begin{equation}
M_{k}^{r,s}=\underset{t=0}{\sum^{k}}\frac{1}{p_{t}^{(r,s)}}\left( \hat{M}%
_{t}^{r,s}-\hat{M}_{t+1}^{r,s}\right) .  \label{81}
\end{equation}

\noindent \qquad We now process $\mathcal{L}_{(R)}$, by separating out the $%
\hat{0}$ components in the sums. This leads to the different contributions:

\noindent I)$\quad u$, $v>r+1$:%
\begin{eqnarray}
&&\mathcal{L}_{(R)}^{I}=\frac{1}{8}\underset{r=0}{\sum^{R}}\underset{u,v=r+2}%
{\sum^{R+1}}\sum_{\left\{ \hat{\gamma},\hat{\mu},\hat{\nu}\right\} }\left(
1+\delta _{u,v}\delta _{\hat{\mu}_{u-1}^{\prime },\nu _{u-1}^{\prime
}}\right) \left( \frac{n_{r}-1}{n_{r}}\right) \hat{M}_{u,v}^{r,r}\times
\label{82} \\
&&\times \left. 
\begin{array}{l}
\left[ \hat{\gamma}_{0}\ldots \hat{\gamma}_{r-1}\left( 
\begin{array}{c}
\hat{\mu}_{r} \\ 
\hat{\gamma}_{r}-\hat{\mu}_{r}%
\end{array}%
\right) \left. 
\begin{array}{c}
\hat{\mu}_{r+1}\ldots \hat{\mu}_{u-1}^{\prime }\hat{0}_{u}\ldots \hat{0}%
_{v-1}\hat{0}_{v}\ldots \hat{0}_{R} \\ 
\hat{\nu}_{r+1}\ldots \hat{\nu}_{u-1}\hat{\nu}_{u}\ldots \hat{\nu}%
_{v-1}^{\prime }\hat{0}_{v}\ldots \hat{0}_{R}%
\end{array}%
\right. \right] _{SN} \\ 
\left[ \hat{\gamma}_{0}\ldots \hat{\gamma}_{r-1}\left( 
\begin{array}{c}
\hat{\mu}_{r} \\ 
\hat{\gamma}_{r}-\hat{\mu}_{r}%
\end{array}%
\right) \left. 
\begin{array}{c}
\hat{\mu}_{r+1}\ldots \hat{\mu}_{u-1}^{\prime }\hat{0}_{u}\ldots \hat{0}%
_{v-1}\hat{0}_{v}\ldots \hat{0}_{R} \\ 
\hat{\nu}_{r+1}\ldots \hat{\nu}_{u-1}\hat{\nu}_{u}\ldots \hat{\nu}%
_{v-1}^{\prime }\hat{0}_{v}\ldots \hat{0}_{R}%
\end{array}%
\right. \right] _{SN}^{\ast }%
\end{array}%
\right.  \notag
\end{eqnarray}

\noindent with field symmetrization at the \textit{marker},%
\begin{equation}
\left( 
\begin{array}{c}
\hat{\mu}_{r} \\ 
\hat{\gamma}_{r}-\hat{\mu}_{r}%
\end{array}%
\right) _{S}=\frac{1}{2}\left[ \left( 
\begin{array}{c}
\hat{\mu}_{r} \\ 
\hat{\gamma}_{r}-\hat{\mu}_{r}%
\end{array}%
\right) _{N}+\left( 
\begin{array}{c}
\hat{\gamma}_{r}-\hat{\mu}_{r} \\ 
\hat{\mu}_{r}%
\end{array}%
\right) _{N}\right]  \label{83}
\end{equation}

\noindent and normalization,%
\begin{equation}
\left( 
\begin{array}{c}
\hat{\mu}_{r} \\ 
\hat{\gamma}_{r}-\hat{\mu}_{r}%
\end{array}%
\right) _{S}=\sqrt{\frac{1}{2}\left( 1+\delta _{u,v}\delta _{\hat{\mu}%
_{u-1}^{\prime },\nu _{u-1}^{\prime }}\right) }\left( 
\begin{array}{c}
\hat{\mu}_{r} \\ 
\hat{\gamma}_{r}-\hat{\mu}_{r}%
\end{array}%
\right) _{SN}.  \label{84}
\end{equation}

\noindent In (82), $\hat{M}_{u,v}^{r,r}$ is the mass double RFT [21],%
\begin{equation}
\hat{M}_{u,v}^{r,r}=\underset{k=u}{\sum^{R+1}}\underset{l=v}{\sum^{R+1}}%
p_{k}p_{l}\left(
M_{k,l}^{r,r}-M_{k-1,l}^{r,r}-M_{k,l-1}^{r,r}+M_{k-1,l-1}^{r,r}\right)
\label{85}
\end{equation}

\noindent the inverse double transform being given by%
\begin{eqnarray}
&&M_{k,l}^{r,r}-M_{k,r}^{r,r}-M_{r,l}^{r,r}+M_{r,r}^{r,r}=  \label{86} \\
&&\quad =\underset{u=r+1}{\sum^{k}}\underset{v=r+1}{\sum^{l}}\frac{1}{p_{u}}%
\frac{1}{p_{v}}\left( \hat{M}_{u+1,v+1}^{r,r}-\hat{M}_{u,v+1}^{r,r}-\hat{M}%
_{u+1,v}^{r,r}+\hat{M}_{u,v}^{r,r}\right) .  \notag
\end{eqnarray}%
\noindent

\noindent II)$\quad u=r+1$, $v>r+1$ (or $v=r+1$, $u>r+1$):

Here one has to separate the $\hat{0}_{r}$ component in the \textit{marker}.

We define the new field, with $\hat{\mu}_{r}^{\prime }\neq \hat{0}_{r}$,%
\begin{equation}
\left\{ 
\begin{array}{c}
\hat{\mu}_{r}^{\prime } \\ 
\hat{\gamma}_{r}-\hat{\mu}_{r}^{\prime }%
\end{array}%
\right\} =\left( 
\begin{array}{c}
\hat{\mu}_{r}^{\prime } \\ 
\hat{\gamma}_{r}-\hat{\mu}_{r}^{\prime }%
\end{array}%
\right) +\frac{1}{n_{r}-1}\left( 
\begin{array}{c}
\hat{0}_{r} \\ 
\hat{\gamma}_{r}%
\end{array}%
\right)  \label{87}
\end{equation}

\noindent which, from (68), has the property%
\begin{equation}
\underset{\hat{\mu}_{r}^{\prime }}{\sum }\left\{ 
\begin{array}{c}
\hat{\mu}_{r}^{\prime } \\ 
\hat{\gamma}_{r}-\hat{\mu}_{r}^{\prime }%
\end{array}%
\right\} =0.  \label{88}
\end{equation}%
Introducing this field, with symmetrization%
\begin{equation}
\left\{ 
\begin{array}{c}
\hat{\mu}_{r}^{\prime } \\ 
\hat{\gamma}_{r}-\hat{\mu}_{r}^{\prime }%
\end{array}%
\right\} _{S}=\frac{1}{2}\left[ \left\{ 
\begin{array}{c}
\hat{\mu}_{r}^{\prime } \\ 
\hat{\gamma}_{r}-\hat{\mu}_{r}^{\prime }%
\end{array}%
\right\} +\left\{ 
\begin{array}{c}
\hat{\gamma}_{r}-\hat{\mu}_{r}^{\prime } \\ 
\hat{\mu}_{r}^{\prime }%
\end{array}%
\right\} \right]  \label{89}
\end{equation}%
and normalization%
\begin{equation}
\left\{ 
\begin{array}{c}
\hat{\mu}_{r}^{\prime } \\ 
\hat{\gamma}_{r}-\hat{\mu}_{r}^{\prime }%
\end{array}%
\right\} _{S}=\sqrt{\frac{1}{2}\left( \frac{n_{r}-2}{n_{r}-1}\right) }%
\left\{ 
\begin{array}{c}
\hat{\mu}_{r}^{\prime } \\ 
\hat{\gamma}_{r}-\hat{\mu}_{r}^{\prime }%
\end{array}%
\right\} _{SN}  \label{90}
\end{equation}

\noindent and using the relation%
\begin{eqnarray}
\underset{\hat{\mu}_{r}^{\prime }\neq \hat{0}_{r}}{\sum }\left\{ 
\begin{array}{c}
\hat{\mu}_{r}^{\prime } \\ 
\hat{\gamma}_{r}-\hat{\mu}_{r}^{\prime }%
\end{array}%
\right\} \left\{ 
\begin{array}{c}
\hat{\mu}_{r}^{\prime } \\ 
\hat{\gamma}_{r}-\hat{\mu}_{r}^{\prime }%
\end{array}%
\right\} ^{\ast } &=&\underset{\hat{\mu}_{r}}{\sum }\left( 
\begin{array}{c}
\hat{\mu}_{r} \\ 
\hat{\gamma}_{r}-\hat{\mu}_{r}%
\end{array}%
\right) \left( 
\begin{array}{c}
\hat{\mu}_{r} \\ 
\hat{\gamma}_{r}-\hat{\mu}_{r}%
\end{array}%
\right) ^{\ast }  \notag \\
&&-\left( 
\begin{array}{c}
\hat{0}_{r} \\ 
\hat{\gamma}_{r}%
\end{array}%
\right) _{N}\left( 
\begin{array}{c}
\hat{0}_{r} \\ 
\hat{\gamma}_{r}%
\end{array}%
\right) _{N}^{\ast },  \label{91}
\end{eqnarray}%
one obtains two contributions:%
\begin{eqnarray}
\mathcal{L}_{(R)}^{II} &=&\frac{1}{8}\underset{r=0}{\sum^{R}}\underset{v=r+2}%
{\sum^{R+1}}\sum_{\left\{ \hat{\gamma},\hat{\mu}^{\prime },\hat{\nu}\right\}
}\left( \frac{n_{r}-2}{n_{r}-1}\right) \hat{M}_{r+1,v}^{r,r}\times 
\label{92} \\
&&\times \left. 
\begin{array}{l}
\left[ \hat{\gamma}_{0}\ldots \hat{\gamma}_{r-1}\left\{ 
\begin{array}{c}
\hat{\mu}_{r}^{\prime } \\ 
\hat{\gamma}_{r}-\hat{\mu}_{r}^{\prime }%
\end{array}%
\right\} \left. 
\begin{array}{c}
\hat{0}_{r+1}\ldots \hat{0}_{v-1}\hat{0}_{v}\ldots \hat{0}_{R} \\ 
\hat{\nu}_{r+1}\ldots \hat{\nu}_{v-1}^{\prime }\hat{0}_{v}\ldots \hat{0}_{R}%
\end{array}%
\right. \right] _{SN} \\ 
\left[ \hat{\gamma}_{0}\ldots \hat{\gamma}_{r-1}\left\{ 
\begin{array}{c}
\hat{\mu}_{r}^{\prime } \\ 
\hat{\gamma}_{r}-\hat{\mu}_{r}^{\prime }%
\end{array}%
\right\} \left. 
\begin{array}{c}
\hat{0}_{r+1}\ldots \hat{0}_{v-1}\hat{0}_{v}\ldots \hat{0}_{R} \\ 
\hat{\nu}_{r+1}\ldots \hat{\nu}_{v-1}^{\prime }\hat{0}_{v}\ldots \hat{0}_{R}%
\end{array}%
\right. \right] _{SN}^{\ast }%
\end{array}%
\right.   \notag
\end{eqnarray}

\noindent and%
\begin{eqnarray}
\mathcal{L}_{(LA)}^{(1)} &=&\frac{1}{8}\underset{r=0}{\sum^{R}}\underset{%
v=r+2}{\sum^{R+1}}\sum_{\left\{ \hat{\gamma}\right\} }\hat{M}%
_{r+1,v}^{r,r}\times  \label{93} \\
&&\times \left. 
\begin{array}{l}
\left[ \hat{\gamma}_{0}\ldots \hat{\gamma}_{r-1}\left( 
\begin{array}{c}
\hat{0}_{r} \\ 
\hat{\gamma}_{r}%
\end{array}%
\right) \left. 
\begin{array}{c}
\hat{0}_{r+1}\ldots \hat{0}_{v-1}\hat{0}_{v}\ldots \hat{0}_{R} \\ 
\hat{\gamma}_{r+1}\ldots \hat{\gamma}_{v-1}^{\prime }\hat{0}_{v}\ldots \hat{0%
}_{R}%
\end{array}%
\right. \right] _{SN} \\ 
\left[ \hat{\gamma}_{0}\ldots \hat{\gamma}_{r-1}\left( 
\begin{array}{c}
\hat{0}_{r} \\ 
\hat{\gamma}_{r}%
\end{array}%
\right) \left. 
\begin{array}{c}
\hat{0}_{r+1}\ldots \hat{0}_{v-1}\hat{0}_{v}\ldots \hat{0}_{R} \\ 
\hat{\gamma}_{r+1}\ldots \hat{\gamma}_{v-1}^{\prime }\hat{0}_{v}\ldots \hat{0%
}_{R}%
\end{array}%
\right. \right] _{SN}^{\ast }%
\end{array}%
\right. .  \notag
\end{eqnarray}

\noindent III)$\quad u=r+1$, $v=r+1$:

Here again one has to separate the $\hat{0}_{r}$ components in the \textit{%
marker}, having now two situations:

A)$\quad \hat{\gamma}_{r}=\hat{0}_{r}$:

We define the new field, with $\hat{\mu}_{r}^{\prime }\neq \hat{0}_{r}$,%
\begin{equation}
\left\{ 
\begin{array}{c}
\hat{\mu}_{r}^{\prime } \\ 
-\hat{\mu}_{r}^{\prime }%
\end{array}%
\right\} =\left( 
\begin{array}{c}
\hat{\mu}_{r}^{\prime } \\ 
-\hat{\mu}_{r}^{\prime }%
\end{array}%
\right) +\frac{1}{n_{r}-1}\left( 
\begin{array}{c}
\hat{0}_{r} \\ 
\hat{0}_{r}%
\end{array}%
\right)  \label{94}
\end{equation}

\noindent which, from (68), has the property%
\begin{equation}
\underset{\hat{\mu}_{r}^{\prime }}{\sum }\left\{ 
\begin{array}{c}
\hat{\mu}_{r}^{\prime } \\ 
-\hat{\mu}_{r}^{\prime }%
\end{array}%
\right\} =0.  \label{95}
\end{equation}%
Introducing this field, with symmetrization%
\begin{equation}
\left\{ 
\begin{array}{c}
\hat{\mu}_{r}^{\prime } \\ 
-\hat{\mu}_{r}^{\prime }%
\end{array}%
\right\} _{S}=\frac{1}{2}\left[ \left\{ 
\begin{array}{c}
\hat{\mu}_{r}^{\prime } \\ 
-\hat{\mu}_{r}^{\prime }%
\end{array}%
\right\} +\left\{ 
\begin{array}{c}
-\hat{\mu}_{r}^{\prime } \\ 
\hat{\mu}_{r}^{\prime }%
\end{array}%
\right\} \right]  \label{96}
\end{equation}%
and normalization 
\begin{equation}
\left\{ 
\begin{array}{c}
\hat{\mu}_{r}^{\prime } \\ 
-\hat{\mu}_{r}^{\prime }%
\end{array}%
\right\} _{S}=\sqrt{\frac{1}{2}\left( \frac{n_{r}-3}{n_{r}-1}\right) }%
\left\{ 
\begin{array}{c}
\hat{\mu}_{r}^{\prime } \\ 
-\hat{\mu}_{r}^{\prime }%
\end{array}%
\right\} _{SN}  \label{97}
\end{equation}%
and using the relation%
\begin{eqnarray}
\underset{\hat{\mu}_{r}^{\prime }\neq \hat{0}_{r}}{\sum }\left\{ 
\begin{array}{c}
\hat{\mu}_{r}^{\prime } \\ 
-\hat{\mu}_{r}^{\prime }%
\end{array}%
\right\} _{S}\left\{ 
\begin{array}{c}
\hat{\mu}_{r}^{\prime } \\ 
-\hat{\mu}_{r}^{\prime }%
\end{array}%
\right\} _{S}^{\ast } &=&\underset{\hat{\mu}_{r}}{\sum }\left( 
\begin{array}{c}
\hat{\mu}_{r} \\ 
-\hat{\mu}_{r}%
\end{array}%
\right) _{S}\left( 
\begin{array}{c}
\hat{\mu}_{r} \\ 
-\hat{\mu}_{r}%
\end{array}%
\right) _{S}^{\ast }  \notag \\
&&-\left( 
\begin{array}{c}
\hat{0}_{r} \\ 
\hat{0}_{r}%
\end{array}%
\right) _{SN}\left( 
\begin{array}{c}
\hat{0}_{r} \\ 
\hat{0}_{r}%
\end{array}%
\right) _{SN}^{\ast }  \label{98}
\end{eqnarray}

\noindent one obtains two contributions:%
\begin{eqnarray}
\mathcal{L}_{(R)}^{III} &=&\frac{1}{8}\underset{r=0}{\sum^{R}}\sum_{\left\{ 
\hat{\gamma},\hat{\mu}^{\prime }\right\} }\left( \frac{n_{r}-3}{n_{r}-1}%
\right) \hat{M}_{r+1,r+1}^{r,r}\times  \label{99} \\
&&\times \left. 
\begin{array}{l}
\left[ \hat{\gamma}_{0}\ldots \hat{\gamma}_{r-1}\left\{ 
\begin{array}{c}
\hat{\mu}_{r}^{\prime } \\ 
-\hat{\mu}_{r}^{\prime }%
\end{array}%
\right\} \left. 
\begin{array}{c}
\hat{0}_{r+1}\ldots \hat{0}_{R} \\ 
\hat{0}_{r+1}\ldots \hat{0}_{R}%
\end{array}%
\right. \right] _{SN} \\ 
\left[ \hat{\gamma}_{0}\ldots \hat{\gamma}_{r-1}\left\{ 
\begin{array}{c}
\hat{\mu}_{r}^{\prime } \\ 
-\hat{\mu}_{r}^{\prime }%
\end{array}%
\right\} \left. 
\begin{array}{c}
\hat{0}_{r+1}\ldots \hat{0}_{R} \\ 
\hat{0}_{r+1}\ldots \hat{0}_{R}%
\end{array}%
\right. \right] _{SN}^{\ast }%
\end{array}%
\right.  \notag
\end{eqnarray}%
and%
\begin{eqnarray}
\mathcal{L}_{(LA)}^{(2)} &=&\frac{1}{4}\underset{r=0}{\sum^{R}}\underset{k=0}%
{\sum^{r}}\sum_{\left\{ \hat{\gamma}\right\} }\hat{M}_{r+1,r+1}^{r,r}\times
\label{100} \\
&&\times \left. 
\begin{array}{l}
\left[ \hat{\gamma}_{0}\ldots \hat{\gamma}_{k-1}^{\prime }\hat{0}_{k}\ldots 
\hat{0}_{r-1}\left( 
\begin{array}{c}
\hat{0}_{r} \\ 
\hat{0}_{r}%
\end{array}%
\right) \left. 
\begin{array}{c}
\hat{0}_{r+1}\ldots \hat{0}_{R} \\ 
\hat{0}_{r+1}\ldots \hat{0}_{R}%
\end{array}%
\right. \right] _{SN} \\ 
\left[ \hat{\gamma}_{0}\ldots \hat{\gamma}_{k-1}^{\prime }\hat{0}_{k}\ldots 
\hat{0}_{r-1}\left( 
\begin{array}{c}
\hat{0}_{r} \\ 
\hat{0}_{r}%
\end{array}%
\right) \left. 
\begin{array}{c}
\hat{0}_{r+1}\ldots \hat{0}_{R} \\ 
\hat{0}_{r+1}\ldots \hat{0}_{R}%
\end{array}%
\right. \right] _{SN}^{\ast }%
\end{array}%
\right.  \notag
\end{eqnarray}

\noindent after successively splitting $\hat{\gamma}_{k}$ into $\hat{0}_{k}$
and $\hat{\gamma}_{k}^{\prime }$.

B)$\quad \hat{\gamma}_{r}^{\prime }\neq 0:$

We define the new field, with $\hat{\mu}_{r}^{\prime \prime }\neq \hat{0}%
_{r},\hat{\gamma}_{r}^{\prime }$,%
\begin{equation}
\left\{ 
\begin{array}{c}
\hat{\mu}_{r}^{\prime \prime } \\ 
\hat{\gamma}_{r}^{\prime }-\hat{\mu}_{r}^{\prime \prime }%
\end{array}%
\right\} =\left( 
\begin{array}{c}
\hat{\mu}_{r}^{\prime \prime } \\ 
\hat{\gamma}_{r}^{\prime }-\hat{\mu}_{r}^{\prime \prime }%
\end{array}%
\right) +\frac{1}{n_{r}-2}\left[ \left( 
\begin{array}{c}
\hat{0}_{r} \\ 
\hat{\gamma}_{r}^{\prime }%
\end{array}%
\right) +\left( 
\begin{array}{c}
\hat{\gamma}_{r}^{\prime } \\ 
\hat{0}_{r}%
\end{array}%
\right) \right]  \label{101}
\end{equation}

\noindent which, from (68), has the property%
\begin{equation}
\underset{\hat{\mu}_{r}^{\prime \prime }}{\sum }\left\{ 
\begin{array}{c}
\hat{\mu}_{r}^{\prime \prime } \\ 
\hat{\gamma}_{r}^{\prime }-\hat{\mu}_{r}^{\prime \prime }%
\end{array}%
\right\} =0.  \label{102}
\end{equation}%
Introducing this field, with symmetrization%
\begin{equation}
\left\{ 
\begin{array}{c}
\hat{\mu}_{r}^{\prime \prime } \\ 
\hat{\gamma}_{r}^{\prime }-\hat{\mu}_{r}^{\prime \prime }%
\end{array}%
\right\} _{S}=\frac{1}{2}\left[ \left\{ 
\begin{array}{c}
\hat{\mu}_{r}^{\prime \prime } \\ 
\hat{\gamma}_{r}^{\prime }-\hat{\mu}_{r}^{\prime \prime }%
\end{array}%
\right\} +\left\{ 
\begin{array}{c}
\hat{\gamma}_{r}^{\prime }-\hat{\mu}_{r}^{\prime \prime } \\ 
\hat{\mu}_{r}^{\prime \prime }%
\end{array}%
\right\} \right]  \label{103}
\end{equation}%
and normalization%
\begin{equation}
\left\{ 
\begin{array}{c}
\hat{\mu}_{r}^{\prime \prime } \\ 
\hat{\gamma}_{r}^{\prime }-\hat{\mu}_{r}^{\prime \prime }%
\end{array}%
\right\} _{S}=\sqrt{\frac{1}{2}\left( \delta _{\hat{\mu}_{r}^{\prime \prime
},\hat{\gamma}_{r}^{\prime }-\hat{\mu}_{r}^{\prime \prime }}+\frac{n_{r}-4}{%
n_{r}-2}\right) }\left\{ 
\begin{array}{c}
\hat{\mu}_{r}^{\prime \prime } \\ 
\hat{\gamma}_{r}^{\prime }-\hat{\mu}_{r}^{\prime \prime }%
\end{array}%
\right\} _{SN}  \label{104}
\end{equation}%
and using the relation%
\begin{eqnarray}
\underset{\hat{\mu}_{r}^{\prime \prime }\neq \hat{0}_{r},\hat{\gamma}%
_{r}^{\prime }}{\sum }\left\{ 
\begin{array}{c}
\hat{\mu}_{r}^{\prime \prime } \\ 
\hat{\gamma}_{r}^{\prime }-\hat{\mu}_{r}^{\prime \prime }%
\end{array}%
\right\} _{S}\left\{ 
\begin{array}{c}
\hat{\mu}_{r}^{\prime \prime } \\ 
\hat{\gamma}_{r}^{\prime }-\hat{\mu}_{r}^{\prime \prime }%
\end{array}%
\right\} _{S}^{\ast } &=&\underset{\hat{\mu}_{r}}{\sum }\left( 
\begin{array}{c}
\hat{\mu}_{r} \\ 
\hat{\gamma}_{r}^{\prime }-\hat{\mu}_{r}%
\end{array}%
\right) _{S}\left( 
\begin{array}{c}
\hat{\mu}_{r} \\ 
\hat{\gamma}_{r}^{\prime }-\hat{\mu}_{r}%
\end{array}%
\right) _{S}^{\ast }  \notag \\
&&-\left( 
\begin{array}{c}
\hat{0}_{r} \\ 
\hat{\gamma}_{r}^{\prime }%
\end{array}%
\right) _{SN}\left( 
\begin{array}{c}
\hat{0}_{r} \\ 
\hat{\gamma}_{r}^{\prime }%
\end{array}%
\right) _{SN}^{\ast }  \label{105}
\end{eqnarray}

\noindent one obtains two contributions:%
\begin{eqnarray}
\mathcal{L}_{(R)}^{IV} &=&\frac{1}{8}\underset{r=0}{\sum^{R}}\sum_{\left\{ 
\hat{\gamma},\hat{\mu}^{\prime \prime }\right\} }\left( \delta _{\hat{\mu}%
_{r}^{\prime \prime },\hat{\gamma}_{r}^{\prime }-\hat{\mu}_{r}^{\prime
\prime }}+\frac{n_{r}-4}{n_{r}-2}\right) \hat{M}_{r+1,r+1}^{r,r}\times
\label{106} \\
&&\times \left. 
\begin{array}{l}
\left[ \hat{\gamma}_{0}\ldots \hat{\gamma}_{r-1}\left\{ 
\begin{array}{c}
\hat{\mu}_{r}^{\prime \prime } \\ 
\hat{\gamma}_{r}^{\prime }-\hat{\mu}_{r}^{\prime \prime }%
\end{array}%
\right\} \left. 
\begin{array}{c}
\hat{0}_{r+1}\ldots \hat{0}_{R} \\ 
\hat{0}_{r+1}\ldots \hat{0}_{R}%
\end{array}%
\right. \right] _{SN} \\ 
\left[ \hat{\gamma}_{0}\ldots \hat{\gamma}_{r-1}\left\{ 
\begin{array}{c}
\hat{\mu}_{r}^{\prime \prime } \\ 
\hat{\gamma}_{r}^{\prime }-\hat{\mu}_{r}^{\prime \prime }%
\end{array}%
\right\} \left. 
\begin{array}{c}
\hat{0}_{r+1}\ldots \hat{0}_{R} \\ 
\hat{0}_{r+1}\ldots \hat{0}_{R}%
\end{array}%
\right. \right] _{SN}^{\ast }%
\end{array}%
\right.  \notag
\end{eqnarray}%
and%
\begin{eqnarray}
\mathcal{L}_{(LA)}^{(3)} &=&\frac{1}{4}\underset{r=0}{\sum^{R}}\sum_{\left\{ 
\hat{\gamma}\right\} }\hat{M}_{r+1,r+1}^{r,r}\times  \label{107} \\
&&\times \left. 
\begin{array}{l}
\left[ \hat{\gamma}_{0}\ldots \hat{\gamma}_{r-1}\left( 
\begin{array}{c}
\hat{0}_{r} \\ 
\hat{\gamma}_{r}^{\prime }%
\end{array}%
\right) \left. 
\begin{array}{c}
\hat{0}_{r+1}\ldots \hat{0}_{R} \\ 
\hat{0}_{r+1}\ldots \hat{0}_{R}%
\end{array}%
\right. \right] _{SN} \\ 
\left[ \hat{\gamma}_{0}\ldots \hat{\gamma}_{r-1}\left( 
\begin{array}{c}
\hat{0}_{r} \\ 
\hat{\gamma}_{r}^{\prime }%
\end{array}%
\right) \left. 
\begin{array}{c}
\hat{0}_{r+1}\ldots \hat{0}_{R} \\ 
\hat{0}_{r+1}\ldots \hat{0}_{R}%
\end{array}%
\right. \right] _{SN}^{\ast }%
\end{array}%
\right. .  \notag
\end{eqnarray}

Thus, we observe that in the replicon geometry there is a
longitudinal-anomalous contribution, given by the components in (93), (100),
(107), which will be used later to calculate the complete
longitudinal-anomalous contribution.

Putting together (82), (92), (99) and (106) one obtains the complete
replicon contribution, $\mathcal{L}_{\mathcal{R}}=\mathcal{L}_{(R)}^{I}+%
\mathcal{L}_{(R)}^{II}+\mathcal{L}_{(R)}^{III}+\mathcal{L}_{(R)}^{IV}$,%
\begin{equation}
\mathcal{L}_{\mathcal{R}}=\frac{1}{2}\underset{r=0}{\sum^{R}}\left\{ 
\underset{u,v=r+2}{\sum^{R+1}}\sum_{\left\{ \hat{\gamma},\hat{\mu},\hat{\nu}%
\right\} }\frac{1}{2}\hat{M}_{u,v}^{r,r}\left\vert ^{\mathcal{R}}\Phi
_{u,v}^{r}\left( 
\begin{array}{c}
\hat{\mu}_{r} \\ 
\hat{\gamma}_{r}-\hat{\mu}_{r}%
\end{array}%
\right) \right\vert ^{2}\right.  \label{108}
\end{equation}%
\begin{equation*}
+\underset{v=r+2}{\sum^{R+1}}\sum_{\left\{ \hat{\gamma},\hat{\mu}^{\prime },%
\hat{\nu}\right\} }\frac{1}{2}\hat{M}_{r+1,v}^{r,r}\left\vert ^{\mathcal{R}%
}\Phi _{r+1,v}^{r}\left( 
\begin{array}{c}
\hat{\mu}_{r}^{\prime } \\ 
\hat{\gamma}_{r}-\hat{\mu}_{r}^{\prime }%
\end{array}%
\right) \right\vert ^{2}
\end{equation*}%
\begin{equation*}
+\underset{u=r+2}{\sum^{R+1}}\sum_{\left\{ \hat{\gamma},\hat{\mu}^{\prime },%
\hat{\nu}\right\} }\frac{1}{2}\hat{M}_{u,r+1}^{r,r}\left\vert ^{\mathcal{R}%
}\Phi _{u,r+1}^{r}\left( 
\begin{array}{c}
\hat{\mu}_{r}^{\prime } \\ 
\hat{\gamma}_{r}-\hat{\mu}_{r}^{\prime }%
\end{array}%
\right) \right\vert ^{2}
\end{equation*}%
\begin{equation*}
+\sum_{\left\{ \hat{\gamma},\hat{\mu}^{\prime }\right\} }\hat{M}%
_{r+1,r+1}^{r,r}{}\left\vert ^{\mathcal{R}}\Phi _{r+1,r+1}^{r}\left( 
\begin{array}{c}
\hat{\mu}_{r}^{\prime } \\ 
-\hat{\mu}_{r}^{\prime }%
\end{array}%
\right) \right\vert ^{2}
\end{equation*}

\begin{equation*}
+\left. \sum_{\left\{ \hat{\gamma},\hat{\mu}^{\prime \prime }\right\} }\hat{M%
}_{r+1,r+1}^{r,r}\left\vert ^{\mathcal{R}}\Phi _{r+1,r+1}^{r}\left( 
\begin{array}{c}
\hat{\mu}_{r}^{\prime \prime } \\ 
\hat{\gamma}_{r}^{\prime }-\hat{\mu}_{r}^{\prime \prime }%
\end{array}%
\right) \right\vert ^{2}\right\}
\end{equation*}

\noindent where $\hat{M}_{u,v}^{r,r}$ is the replicon mass, given by (85),
and $^{\mathcal{R}}\Phi _{u,v}^{r}$ are the replicon fields, defined as: \ 

\noindent $\bullet $ $u$, $v>r+1$:%
\begin{eqnarray}
&&^{\mathcal{R}}\Phi _{u,v}^{r}\left( 
\begin{array}{c}
\hat{\mu}_{r} \\ 
\hat{\gamma}_{r}-\hat{\mu}_{r}%
\end{array}%
\right) =  \label{109} \\
&&\quad =\mathcal{N}_{1}\left[ \hat{\gamma}_{0}\ldots \hat{\gamma}%
_{r-1}\left( 
\begin{array}{c}
\hat{\mu}_{r} \\ 
\hat{\gamma}_{r}-\hat{\mu}_{r}%
\end{array}%
\right) \left. 
\begin{array}{c}
\hat{\mu}_{r+1}\ldots \hat{\mu}_{u-1}^{\prime }\hat{0}_{u}\ldots \hat{0}%
_{v-1}\hat{0}_{v}\ldots \hat{0}_{R} \\ 
\hat{\nu}_{r+1}\ldots \hat{\nu}_{u-1}\hat{\nu}_{u}\ldots \hat{\nu}%
_{v-1}^{\prime }\hat{0}_{v}\ldots \hat{0}_{R}%
\end{array}%
\right. \right] _{SN}  \notag
\end{eqnarray}%
\noindent

\noindent with $\mathcal{N}_{1}=\sqrt{\frac{1}{2}\left( 1+\delta
_{u,v}\delta _{\hat{\mu}_{u-1}^{\prime },\hat{\nu}_{u-1}^{\prime }}\right)
\left( \frac{n_{r}-1}{n_{r}}\right) }$, having the property%
\begin{equation}
\underset{\hat{\mu}_{r}}{\sum }\,^{\mathcal{R}}\Phi _{u,v}^{r}\left( 
\begin{array}{c}
\hat{\mu}_{r} \\ 
\hat{\gamma}_{r}-\hat{\mu}_{r}%
\end{array}%
\right) =0  \label{110}
\end{equation}

\noindent and multiplicity%
\begin{equation}
\mu (r;u,v)=\frac{1}{2}p_{0}\left( p_{r}-p_{r+1}\right) \left( \frac{1}{p_{u}%
}-\frac{1}{p_{u-1}}\right) \left( \frac{1}{p_{v}}-\frac{1}{p_{v-1}}\right) ;
\label{111}
\end{equation}

\noindent $\bullet $ $u=r+1$, $v>r+1$ (or $v=r+1$, $u>r+1$):%
\begin{eqnarray}
&&^{\mathcal{R}}\Phi _{r+1,v}^{r}\left( 
\begin{array}{c}
\hat{\mu}_{r}^{\prime } \\ 
\hat{\gamma}_{r}-\hat{\mu}_{r}^{\prime }%
\end{array}%
\right) =  \label{112} \\
&&\quad =\mathcal{N}_{2}\left[ \hat{\gamma}_{0}\ldots \hat{\gamma}%
_{r-1}\left\{ 
\begin{array}{c}
\hat{\mu}_{r}^{\prime } \\ 
\hat{\gamma}_{r}-\hat{\mu}_{r}^{\prime }%
\end{array}%
\right\} \left. 
\begin{array}{c}
\hat{0}_{r+1}\ldots \hat{0}_{v-1}\hat{0}_{v}\ldots \hat{0}_{R} \\ 
\hat{\nu}_{r+1}\ldots \hat{\nu}_{v-1}^{\prime }\hat{0}_{v}\ldots \hat{0}_{R}%
\end{array}%
\right. \right] _{SN}  \notag
\end{eqnarray}

\noindent with $\mathcal{N}_{2}=\sqrt{\frac{1}{2}\left( \frac{n_{r}-2}{%
n_{r}-1}\right) }$, having the property%
\begin{equation}
\underset{\hat{\mu}_{r}^{\prime }}{\sum }\,^{\mathcal{R}}\Phi
_{r+1,v}^{r}\left( 
\begin{array}{c}
\hat{\mu}_{r}^{\prime } \\ 
\hat{\gamma}_{r}-\hat{\mu}_{r}^{\prime }%
\end{array}%
\right) =0  \label{113}
\end{equation}

\noindent and multiplicity%
\begin{equation}
\mu (r;r+1,v)=\frac{1}{2}p_{0}\left( \frac{p_{r}}{p_{r+1}}-2\right) \left( 
\frac{1}{p_{v}}-\frac{1}{p_{v-1}}\right) ;  \label{114}
\end{equation}

\noindent $\bullet $ $u=r+1$, $v=r+1$:%
\begin{equation}
^{\mathcal{R}}\Phi _{r+1,r+1}^{r}\left( 
\begin{array}{c}
\hat{\mu}_{r}^{\prime } \\ 
-\hat{\mu}_{r}^{\prime }%
\end{array}%
\right) =\mathcal{N}_{3}\left[ \hat{\gamma}_{0}\ldots \hat{\gamma}%
_{r-1}\left\{ 
\begin{array}{c}
\hat{\mu}_{r}^{\prime } \\ 
-\hat{\mu}_{r}^{\prime }%
\end{array}%
\right\} \left. 
\begin{array}{c}
\hat{0}_{r+1}\ldots \hat{0}_{R} \\ 
\hat{0}_{r+1}\ldots \hat{0}_{R}%
\end{array}%
\right. \right] _{SN}  \label{115}
\end{equation}

\noindent with $\mathcal{N}_{3}=\sqrt{\frac{1}{4}\left( \frac{n_{r}-3}{%
n_{r}-1}\right) }$, having the property%
\begin{equation}
\underset{\hat{\mu}_{r}^{\prime }}{\sum }\,^{\mathcal{R}}\Phi
_{r+1,r+1}^{r}\left( 
\begin{array}{c}
\hat{\mu}_{r}^{\prime } \\ 
-\hat{\mu}_{r}^{\prime }%
\end{array}%
\right) =0  \label{116}
\end{equation}%
and multiplicity%
\begin{equation}
\mu _{1}(r;r+1,r+1)=\frac{1}{2}p_{0}\left( \frac{p_{r}}{p_{r+1}}-3\right) 
\frac{1}{p_{r}};  \label{117}
\end{equation}

\noindent and%
\begin{eqnarray}
&&^{\mathcal{R}}\Phi _{r+1,r+1}^{r}\left( 
\begin{array}{c}
\hat{\mu}_{r}^{\prime \prime } \\ 
\hat{\gamma}_{r}^{\prime }-\hat{\mu}_{r}^{\prime \prime }%
\end{array}%
\right) =  \label{118} \\
&&\qquad \qquad \qquad \quad =\mathcal{N}_{4}\left[ \hat{\gamma}_{0}\ldots 
\hat{\gamma}_{r-1}\left\{ 
\begin{array}{c}
\hat{\mu}_{r}^{\prime \prime } \\ 
\hat{\gamma}_{r}^{\prime }-\hat{\mu}_{r}^{\prime \prime }%
\end{array}%
\right\} \left. 
\begin{array}{c}
\hat{0}_{r+1}\ldots \hat{0}_{R} \\ 
\hat{0}_{r+1}\ldots \hat{0}_{R}%
\end{array}%
\right. \right] _{SN}  \notag
\end{eqnarray}

\noindent with $\mathcal{N}_{4}=\sqrt{\frac{1}{4}\left( \delta _{\hat{\mu}%
_{r}^{\prime \prime },\hat{\gamma}_{r}^{\prime }-\hat{\mu}_{r}^{\prime
\prime }}+\frac{n_{r}-4}{n_{r}-2}\right) }$, having the property%
\begin{equation}
\underset{\hat{\mu}_{r}^{\prime \prime }}{\sum }\,^{\mathcal{R}}\Phi
_{r+1,r+1}^{r}\left( 
\begin{array}{c}
\hat{\mu}_{r}^{\prime \prime } \\ 
\hat{\gamma}_{r}^{\prime }-\hat{\mu}_{r}^{\prime \prime }%
\end{array}%
\right) =0  \label{119}
\end{equation}

\noindent and multiplicity

\begin{equation}
\mu _{2}(r;r+1,r+1)=\frac{1}{2}p_{0}\left( \frac{p_{r}}{p_{r+1}}-3\right)
\left( \frac{1}{p_{r+1}}-\frac{1}{p_{r}}\right) ;  \label{120}
\end{equation}%
\noindent defining, $\mu =\mu _{1}+\mu _{2}$, gives%
\begin{equation}
\mu (r;r+1,r+1)=\frac{1}{2}p_{0}\left( \frac{p_{r}}{p_{r+1}}-3\right) \frac{1%
}{p_{r+1}}.  \label{121}
\end{equation}

The propagators for the replicon fields, obtained from (108), are given by:%
\begin{eqnarray}
^{\mathcal{R}}G_{u,v}^{r}\left( \hat{\mu}_{r},\hat{\eta}_{r};\hat{\gamma}%
_{r},\hat{\lambda}_{r}\right) &=&\left\langle ^{\mathcal{R}}\Phi
_{u,v}^{r}\left( 
\begin{array}{c}
\hat{\mu}_{r} \\ 
\hat{\gamma}_{r}-\hat{\mu}_{r}%
\end{array}%
\right) \text{ }^{\mathcal{R}}\Phi _{u,v}^{r}\left( 
\begin{array}{c}
\hat{\eta}_{r} \\ 
\hat{\lambda}_{r}-\hat{\eta}_{r}%
\end{array}%
\right) ^{\ast }\right\rangle  \notag \\
&=&\delta _{\gamma ,\lambda }\left[ \delta _{\mu ,\eta }-\frac{1}{n_{r}}%
\right] \frac{1}{\hat{M}_{u,v}^{r,r}}  \label{122}
\end{eqnarray}%
\begin{eqnarray}
&&^{\mathcal{R}}G_{r+1,v}^{r}\left( \hat{\mu}_{r}^{\prime },\hat{\eta}%
_{r}^{\prime };\hat{\gamma}_{r},\hat{\lambda}_{r}\right) =  \label{123} \\
&&\qquad \qquad \qquad \qquad =\left\langle ^{\mathcal{R}}\Phi
_{r+1,v}^{r}\left( 
\begin{array}{c}
\hat{\mu}_{r}^{\prime } \\ 
\hat{\gamma}_{r}-\hat{\mu}_{r}^{\prime }%
\end{array}%
\right) \text{ }^{\mathcal{R}}\Phi _{r+1,v}^{r}\left( 
\begin{array}{c}
\hat{\eta}_{r}^{\prime } \\ 
\hat{\lambda}_{r}-\hat{\eta}_{r}^{\prime }%
\end{array}%
\right) ^{\ast }\right\rangle  \notag \\
&&\qquad \qquad \qquad \qquad =\delta _{\gamma ,\lambda }\left[ \delta _{\mu
^{\prime },\eta ^{\prime }}-\frac{1}{n_{r}-1}\right] \frac{1}{\hat{M}%
_{r+1,v}^{r,r}}  \notag
\end{eqnarray}%
\begin{eqnarray}
^{\mathcal{R}}G_{r+1,r+1}^{r}\left( \hat{\mu}_{r}^{\prime },\hat{\eta}%
_{r}^{\prime };\hat{0}_{r}\right) &=&\left\langle ^{\mathcal{R}}\Phi
_{r+1,r+1}^{r}\left( 
\begin{array}{c}
\hat{\mu}_{r}^{\prime } \\ 
-\hat{\mu}_{r}^{\prime }%
\end{array}%
\right) \text{ }^{\mathcal{R}}\Phi _{r+1,r+1}^{r}\left( 
\begin{array}{c}
\hat{\eta}_{r}^{\prime } \\ 
-\hat{\eta}_{r}^{\prime }%
\end{array}%
\right) ^{\ast }\right\rangle  \notag \\
&=&\left[ \frac{1}{2}(\delta _{\mu ^{\prime },\eta ^{\prime }}+\delta _{\mu
^{\prime },-\eta ^{\prime }})-\frac{1}{n_{r}-1}\right] \frac{1}{\hat{M}%
_{r+1,r+1}^{r,r}}  \label{124}
\end{eqnarray}%
\begin{eqnarray}
&&^{\mathcal{R}}G_{r+1,r+1}^{r}\left( \hat{\mu}_{r}^{\prime \prime },\hat{%
\eta}_{r}^{\prime \prime };\hat{\gamma}_{r}^{\prime },\hat{\lambda}%
_{r}^{\prime }\right) =  \label{125} \\
&&\qquad \qquad \qquad =\left\langle ^{\mathcal{R}}\Phi _{r+1,r+1}^{r}\left( 
\begin{array}{c}
\hat{\mu}_{r}^{\prime \prime } \\ 
\hat{\gamma}_{r}^{\prime }-\hat{\mu}_{r}^{\prime \prime }%
\end{array}%
\right) \text{ }^{\mathcal{R}}\Phi _{r+1,r+1}^{r}\left( 
\begin{array}{c}
\hat{\eta}_{r}^{\prime \prime } \\ 
\hat{\lambda}_{r}^{\prime }-\hat{\eta}_{r}^{\prime \prime }%
\end{array}%
\right) ^{\ast }\right\rangle  \notag \\
&&\qquad \qquad \qquad =\delta _{\gamma ^{\prime },\lambda ^{\prime }}\left[ 
\frac{1}{2}\left( \delta _{\mu ^{\prime \prime },\eta ^{\prime \prime
}}+\delta _{\gamma ^{\prime }-\mu ^{\prime \prime },\eta ^{\prime \prime
}}\right) -\frac{1}{n_{r}-2}\right] \frac{1}{\hat{M}_{r+1,r+1}^{r,r}}. 
\notag
\end{eqnarray}

Putting together (71), (72), (73), (93), (100) and (107) one obtains the
complete longitudinal-anomalous contribution, $\mathcal{L}_{\mathcal{LA}}=%
\mathcal{L}_{(LA)}^{I}+\mathcal{L}_{(LA)}^{II}+\mathcal{L}_{(LA)}^{III}+%
\mathcal{L}_{(LA)}^{(1)}+\mathcal{L}_{(LA)}^{(2)}+\mathcal{L}_{(LA)}^{(3)}$,%
\begin{equation}
\mathcal{L}_{\mathcal{LA}}=\frac{1}{2}\underset{r=0}{\sum^{R}}\left\{ 
\underset{v=r+2}{\sum^{R+1}}\sum_{\left\{ \hat{\gamma}\right\} }\frac{1}{2}%
\hat{M}_{r+1,v}^{r,r}\left\vert ^{\mathcal{LA}}\Psi _{v}^{r}\left( 
\begin{array}{c}
\hat{0}_{r} \\ 
\hat{\gamma}_{r}%
\end{array}%
\right) \right\vert ^{2}\right.  \label{126}
\end{equation}%
\begin{equation*}
+\underset{u=r+2}{\sum^{R+1}}\sum_{\left\{ \hat{\gamma}\right\} }\frac{1}{2}%
M_{u,r+1}^{r,r}\left\vert ^{\mathcal{LA}}\Psi _{u}^{r}\left( 
\begin{array}{c}
\hat{0}_{r} \\ 
\hat{\gamma}_{r}%
\end{array}%
\right) \right\vert ^{2}
\end{equation*}%
\begin{equation*}
+\underset{k=0}{\sum^{r}}\sum_{\left\{ \hat{\gamma}\right\} }\hat{M}%
_{r+1,r+1}^{r,r}\left\vert ^{\mathcal{LA}}\Psi _{k}^{r}\left( 
\begin{array}{c}
\hat{0}_{r} \\ 
\hat{0}_{r}%
\end{array}%
\right) \right\vert ^{2}
\end{equation*}%
\begin{equation*}
+\left. \sum_{\left\{ \hat{\gamma}\right\} }\hat{M}_{r+1,r+1}^{r,r}\left%
\vert ^{\mathcal{LA}}\Psi _{r+1}^{r}\left( 
\begin{array}{c}
\hat{0}_{r} \\ 
\hat{\gamma}_{r}^{\prime }%
\end{array}%
\right) \right\vert ^{2}\right\}
\end{equation*}%
\begin{equation*}
+\frac{1}{8}\underset{s=r}{\sum^{R}}\underset{r=t}{\sum^{s}}\underset{t=0}{%
\sum^{r}}\sum_{\left\{ \hat{\gamma}\right\} }{}^{\mathcal{LA}}\Psi
_{t}^{r}\left( 
\begin{array}{c}
\hat{0}_{r} \\ 
\hat{0}_{r}%
\end{array}%
\right) \sqrt{\delta _{r}^{(t-1)}}\hat{M}_{t}^{r,s}\sqrt{\delta _{s}^{(t-1)}}%
\text{ }^{\mathcal{LA}}\Psi _{t}^{s}\left( 
\begin{array}{c}
\hat{0}_{s} \\ 
\hat{0}_{s}%
\end{array}%
\right) ^{\ast }
\end{equation*}%
\begin{equation*}
+\frac{1}{8}\underset{r=s+1}{\sum^{R}}\underset{s=t}{\sum^{r-1}}\underset{t=0%
}{\sum^{s}}\sum_{\left\{ \hat{\gamma}\right\} }{}^{\mathcal{LA}}\Psi
_{t}^{r}\left( 
\begin{array}{c}
\hat{0}_{r} \\ 
\hat{0}_{r}%
\end{array}%
\right) \sqrt{\delta _{r}^{(t-1)}}\hat{M}_{t}^{r,s}\sqrt{\delta _{s}^{(t-1)}}%
\text{ }^{\mathcal{LA}}\Psi _{t}^{s}\left( 
\begin{array}{c}
\hat{0}_{s} \\ 
\hat{0}_{s}%
\end{array}%
\right) ^{\ast }
\end{equation*}%
\begin{equation*}
+\frac{1}{8}\underset{s=t}{\sum^{R}}\underset{t=r+1}{\sum^{s}}\underset{r=0}{%
\sum^{t-1}}\sum_{\left\{ \hat{\gamma}\right\} }{}^{\mathcal{LA}}\Psi
_{t}^{r}\left( 
\begin{array}{c}
\hat{0}_{r} \\ 
\hat{\gamma}_{r}%
\end{array}%
\right) \sqrt{\delta _{r}^{(t-1)}}\hat{M}_{t}^{r,s}\sqrt{\delta _{s}^{(t-1)}}%
\text{ }^{\mathcal{LA}}\Psi _{t}^{s}\left( 
\begin{array}{c}
\hat{0}_{s} \\ 
\hat{0}_{s}%
\end{array}%
\right) ^{\ast }
\end{equation*}%
\begin{equation*}
+\frac{1}{8}\underset{r=t}{\sum^{R}}\underset{t=s+1}{\sum^{r}}\underset{s=0}{%
\sum^{t-1}}\sum_{\left\{ \hat{\gamma}\right\} }{}^{\mathcal{LA}}\Psi
_{t}^{r}\left( 
\begin{array}{c}
\hat{0}_{r} \\ 
\hat{0}_{r}%
\end{array}%
\right) \sqrt{\delta _{r}^{(t-1)}}\hat{M}_{t}^{r,s}\sqrt{\delta _{s}^{(t-1)}}%
\text{ }^{\mathcal{LA}}\Psi _{t}^{s}\left( 
\begin{array}{c}
\hat{0}_{s} \\ 
\hat{\gamma}_{s}%
\end{array}%
\right) ^{\ast }
\end{equation*}%
\begin{equation*}
+\frac{1}{8}\underset{t=s+1}{\sum^{R+1}}\underset{s=r}{\sum^{t-1}}\underset{%
r=0}{\sum^{s}}\sum_{\left\{ \hat{\gamma}\right\} }{}^{\mathcal{LA}}\Psi
_{t}^{r}\left( 
\begin{array}{c}
\hat{0}_{r} \\ 
\hat{\gamma}_{r}%
\end{array}%
\right) \sqrt{\delta _{r}^{(t-1)}}\hat{M}_{t}^{r,s}\sqrt{\delta _{s}^{(t-1)}}%
\text{ }^{\mathcal{LA}}\Psi _{t}^{s}\left( 
\begin{array}{c}
\hat{0}_{s} \\ 
\hat{\gamma}_{s}%
\end{array}%
\right) ^{\ast }
\end{equation*}%
\begin{equation*}
+\frac{1}{8}\underset{t=r+1}{\sum^{R+1}}\underset{r=s+1}{\sum^{t-1}}\underset%
{s=0}{\sum^{r-1}}\sum_{\left\{ \hat{\gamma}\right\} }{}^{\mathcal{LA}}\Psi
_{t}^{r}\left( 
\begin{array}{c}
\hat{0}_{r} \\ 
\hat{\gamma}_{r}%
\end{array}%
\right) \sqrt{\delta _{r}^{(t-1)}}\hat{M}_{t}^{r,s}\sqrt{\delta _{s}^{(t-1)}}%
\text{ }^{\mathcal{LA}}\Psi _{t}^{s}\left( 
\begin{array}{c}
\hat{0}_{s} \\ 
\hat{\gamma}_{s}%
\end{array}%
\right) ^{\ast }
\end{equation*}

\noindent where $\hat{M}_{t}^{r,s}$ and $\hat{M}_{u,v}^{r,r}$ are given by
(79) and (85), respectively, and $^{\mathcal{LA}}\Psi _{t}^{r}$ are the
longitudinal-anomalous fields, defined as:

\noindent $\bullet $ $t<r+1$:%
\begin{equation}
^{\mathcal{LA}}\Psi _{t}^{r}\left( 
\begin{array}{c}
\hat{0}_{r} \\ 
\hat{0}_{r}%
\end{array}%
\right) =\frac{1}{\sqrt{2}}\left[ \hat{\gamma}_{0}\ldots \hat{\gamma}%
_{t-1}^{\prime }\hat{0}_{t}\ldots \hat{0}_{r-1}\left( 
\begin{array}{c}
\hat{0}_{r} \\ 
\hat{0}_{r}%
\end{array}%
\right) \left. 
\begin{array}{c}
\hat{0}_{r+1}\ldots \hat{0}_{R} \\ 
\hat{0}_{r+1}\ldots \hat{0}_{R}%
\end{array}%
\right. \right] _{SN}  \label{127}
\end{equation}

\noindent $\bullet $ $t=r+1$:$\ $%
\begin{equation}
^{\mathcal{LA}}\Psi _{t}^{r}\left( 
\begin{array}{c}
\hat{0}_{r} \\ 
\hat{\gamma}_{r}^{\prime }%
\end{array}%
\right) =\frac{1}{\sqrt{2}}\left[ \hat{\gamma}_{0}\ldots \hat{\gamma}%
_{r-1}\left( 
\begin{array}{c}
\hat{0}_{r} \\ 
\hat{\gamma}_{r}^{\prime }%
\end{array}%
\right) \left. 
\begin{array}{c}
\hat{0}_{r+1}\ldots \hat{0}_{R} \\ 
\hat{0}_{r+1}\ldots \hat{0}_{R}%
\end{array}%
\right. \right] _{SN}  \label{128}
\end{equation}

\noindent $\bullet $ $t>r+1$:%
\begin{equation}
^{\mathcal{LA}}\Psi _{t}^{r}\left( 
\begin{array}{c}
\hat{0}_{r} \\ 
\hat{\gamma}_{r}%
\end{array}%
\right) =\frac{1}{\sqrt{2}}\left[ \hat{\gamma}_{0}\ldots \hat{\gamma}%
_{r-1}\left( 
\begin{array}{c}
\hat{0}_{r} \\ 
\hat{\gamma}_{r}%
\end{array}%
\right) \left. 
\begin{array}{c}
\hat{0}_{r+1}\ldots \hat{0}_{t-1}\hat{0}_{t}\ldots \hat{0}_{R} \\ 
\hat{\gamma}_{r+1}\ldots \hat{\gamma}_{t-1}^{\prime }\hat{0}_{t}\ldots \hat{0%
}_{R}%
\end{array}%
\right. \right] _{SN}  \label{129}
\end{equation}

\noindent with multiplicity%
\begin{eqnarray}
\mu (t) &=&p_{0}\left( \frac{1}{p_{t}}-\frac{1}{p_{t-1}}\right) \qquad
\label{130} \\
\mu (0) &=&1.\quad  \notag
\end{eqnarray}

Equation (126) can be written in the generic form,%
\begin{equation}
\mathcal{L}_{\mathcal{LA}}=\frac{1}{2}\underset{t=0}{\sum^{R+1}}\underset{%
r,s=0}{\sum^{R}}\sum_{\left\{ \hat{\gamma}\right\} }\text{ }^{\mathcal{LA}%
}\Psi _{t}^{r}\left[ \delta _{r,s}^{Kr}\hat{\Lambda}_{t}^{r}+\frac{1}{4}%
\sqrt{\delta _{r}^{(t-1)}}\hat{M}_{t}^{r,s}\sqrt{\delta _{s}^{(t-1)}}\right] 
\text{ }^{\mathcal{LA}}\Psi _{t}^{s\ast }  \label{131}
\end{equation}

\noindent with $\hat{\Lambda}_{t}^{r}$ defined as%
\begin{equation}
\hat{\Lambda}_{t}^{r}=\left\{ 
\begin{array}{l}
\hat{M}_{t,r+1}^{r,r}\qquad t>r+1 \\ 
\hat{M}_{r+1,r+1}^{r,r}\qquad t\leq r+1%
\end{array}%
\right. .  \label{132}
\end{equation}%
The propagators:

\begin{equation}
^{\mathcal{LA}}G_{t}^{r,s}\left( \hat{\gamma}_{r};\hat{\lambda}_{s}\right)
=\left\langle ^{\mathcal{LA}}\Psi _{t}^{r}\left( 
\begin{array}{c}
\hat{0}_{r} \\ 
\hat{\gamma}_{r}%
\end{array}%
\right) \,^{\mathcal{LA}}\Psi _{t}^{s\ast }\left( 
\begin{array}{c}
\hat{0}_{s} \\ 
\hat{\lambda}_{s}%
\end{array}%
\right) \right\rangle  \label{133}
\end{equation}

\noindent are given by the inverse of the matrix%
\begin{equation}
\tilde{M}_{t}^{r,s}=\delta _{r,s}^{Kr}\hat{\Lambda}_{t}^{r}+\frac{1}{4}\sqrt{%
\delta _{r}^{(t-1)}}\hat{M}_{t}^{r,s}\sqrt{\delta _{s}^{(t-1)}}  \label{134}
\end{equation}

\noindent that is, [20, 21]%
\begin{equation}
^{\mathcal{LA}}\hat{G}_{t}^{r,s}=\delta _{r,s}^{Kr}\frac{1}{\hat{\Lambda}%
_{t}^{r}}+\frac{1}{4}\sqrt{\delta _{r}^{(t-1)}}\hat{F}_{t}^{r,s}\sqrt{\delta
_{s}^{(t-1)}}  \label{135}
\end{equation}

\noindent with%
\begin{equation}
\hat{F}_{t}^{r,s}=-\frac{1}{\hat{\Lambda}_{t}^{r}}\hat{M}_{t}^{r,s}\frac{1}{%
\hat{\Lambda}_{t}^{s}}-\frac{1}{\hat{\Lambda}_{t}^{r}}\underset{k=0}{\sum^{R}%
}\hat{M}_{t}^{r,k}\frac{\delta _{k}^{(t-1)}}{4}\hat{F}_{t}^{k,s}.
\label{136}
\end{equation}

\noindent A fully explicit form for the solution of $\hat{F}_{t}^{r,s}$ can
be found in $[18]$.

From (108) and (126) one sees that the Lagrangian, $\mathcal{L}^{(2)}=%
\mathcal{L}_{\mathcal{LA}}+\mathcal{L}_{\mathcal{R}}$, breaks up into a
string of $(R+1)\times (R+1)$ blocks followed by a string of $1\times 1$
\textquotedblright blocks\textquotedblright\ along the diagonal. The $%
(R+1)\times (R+1)$ blocks correspond to the longitudinal--anomalous sector,
they contain the matrix elements $\hat{M}_{t}^{r,s}$ with $r,s=0,\ldots ,R$,
and are labelled by the index $t=0,1,\ldots ,R+1,$ ($t=0$ is the
longitudinal and $t\neq 0$ are the anomalous). The $1\times 1$
\textquotedblright blocks\textquotedblright\ correspond to the replicon
sector, they are the elements $\hat{M}_{u,v}^{r,r}$ with $r=0,\ldots ,R$ and 
$u,v=r+1,\ldots ,R+1$. The total number of longitudinal-anomalous modes is%
\begin{equation}
\mu _{\mathcal{LA}}=\sum_{r=0}^{R}\sum_{t=0}^{R+1}\mu (t)=(R+1)p_{0},
\label{137}
\end{equation}

\noindent and the total number of replicon modes is%
\begin{equation}
\mu _{\mathcal{R}}=\sum_{r=0}^{R}\sum_{u=r+1}^{R+1}\sum_{v=r+1}^{R+1}\mu
(r;u,v)=\frac{1}{2}p_{0}(p_{0}-1)-(R+1)p_{0},  \label{138}
\end{equation}

\noindent so that the total number of modes is%
\begin{equation}
\mu =\mu _{\mathcal{LA}}+\mu _{\mathcal{R}}=\frac{n(n-1)}{2}\text{.}
\label{139}
\end{equation}

We note that for $R=0$, (108) and (126), with (32) and (33), naturally lead
to (28).

One can easily obtain the propagators in the direct replica space, for
general $R$, in terms of their RFT expression, e.g., for%
\begin{equation}
G^{ab,ab}=G_{R+1,R+1}^{r,r}=\left\langle \phi _{ab}^{S}\phi
_{ab}^{S}\right\rangle =\left\langle \phi _{r}^{S}\phi _{r}^{S}\right\rangle
\label{140}
\end{equation}

\noindent with the symmetrized field%
\begin{eqnarray}
\phi _{r}^{S} &=&\left[ a_{0}\ldots a_{r-1}\left. 
\begin{array}{l}
a_{r}\ldots a_{R} \\ 
b_{r}\ldots b_{R}%
\end{array}%
\right. \right] _{S}  \label{141} \\
&=&\frac{1}{2}\left( \left[ a_{0}\ldots a_{r-1}\left. 
\begin{array}{l}
a_{r}\ldots a_{R} \\ 
b_{r}\ldots b_{R}%
\end{array}%
\right. \right] \right. +\left. \left[ a_{0}\ldots a_{r-1}\left. 
\begin{array}{l}
b_{r}\ldots b_{R} \\ 
a_{r}\ldots a_{R}%
\end{array}%
\right. \right] \right)  \notag
\end{eqnarray}

\noindent one finds%
\begin{equation}
G_{R+1,R+1}^{r,r}=\frac{1}{n_{0}\ldots
n_{r-1}n_{r}(n_{r}-1)n_{r+1}^{2}\ldots n_{R}^{2}}\sum_{\substack{ \left\{
a_{i},b_{i}\right\}  \\ a_{r}\neq b_{r}}}\left\langle \phi _{r}^{S}\phi
_{r}^{S}\right\rangle  \label{142}
\end{equation}%
\begin{equation*}
=\frac{1}{p_{0}(p_{r}-p_{r+1})}\sum_{\left\{ \hat{\gamma}_{i},\hat{\mu}_{i},%
\hat{\nu}_{i}\right\} }\left\{ \sum_{u,v=r+2}^{R+1}\right. \,^{\mathcal{R}%
}G_{u,v}^{r}\left( \hat{\mu}_{r},\hat{\mu}_{r};\hat{\gamma}_{r},\hat{\gamma}%
_{r}\right)
\end{equation*}%
\begin{equation*}
+\sum_{u=r+2}^{R+1}\,^{\mathcal{R}}G_{u,r+1}^{r}\left( \hat{\mu}_{r}^{\prime
},\hat{\mu}_{r}^{\prime };\hat{\gamma}_{r},\hat{\gamma}_{r}\right)
+\sum_{v=r+2}^{R+1}\,^{\mathcal{R}}G_{r+1,v}^{r}\left( \hat{\mu}_{r}^{\prime
},\hat{\mu}_{r}^{\prime };\hat{\gamma}_{r},\hat{\gamma}_{r}\right)
\end{equation*}%
\begin{equation*}
+2^{\mathcal{R}}G_{r+1,r+1}^{r}\left( \hat{\mu}_{r}^{\prime },\hat{\mu}%
_{r}^{\prime };\hat{0}_{r}\right) +\,2^{\mathcal{R}}G_{r+1,r+1}^{r}\left( 
\hat{\mu}_{r}^{\prime \prime },\hat{\mu}_{r}^{\prime \prime };\hat{\gamma}%
_{r}^{\prime },\hat{\gamma}_{r}^{\prime }\right)
\end{equation*}%
\begin{equation*}
+8\sum_{t=r+2}^{R+1}\,^{\mathcal{LA}}G_{t}^{r,r}\left( \hat{\gamma}_{r};\hat{%
\gamma}_{r}\right) +\,2^{\mathcal{LA}}G_{r+1}^{r,r}\left( \hat{\gamma}%
_{r}^{\prime };\hat{\gamma}_{r}^{\prime }\right) +2\left. \sum_{t=0}^{r}\,^{%
\mathcal{LA}}G_{t}^{r,r}\left( \hat{0}_{r};\hat{0}_{r}\right) \right\} .
\end{equation*}

\bigskip \bigskip

\noindent {\large 5. Spin Glass Free Energy with Fluctuations}

\bigskip

Here we use the RFT formalism to calculate the contribution of the Gaussian
fluctuations around the Parisi's solution for the free energy of an Ising
spin glass. The spin glass free energy, (2), calculated with the partition
function in (3), can be written as 
\begin{equation}
\overline{F}=F_{mf}+F_{fluct}  \label{143}
\end{equation}%
where{} 
\begin{equation}
F_{mf}=\frac{1}{\beta }\underset{n\rightarrow 0}{\lim }\frac{\mathcal{L}%
^{(0)}}{n}  \label{144}
\end{equation}

\noindent provides the mean field value of the free energy, with $\mathcal{L}%
^{(0)}$ given by (7), and 
\begin{equation}
F_{fluct}=-\frac{1}{\beta }\underset{n\rightarrow 0}{\lim }\frac{\ln \left[ 
\overline{Z^{n}}\right] _{fluct}}{n}  \label{145}
\end{equation}

\noindent provides the contribution of the fluctuations. For fluctuations up
to the quadratic order,%
\begin{equation}
\left[ \overline{Z^{n}}\right] _{fluct}=\int \mathcal{D}\left( ^{\mathcal{LA}%
}\Psi \right) \mathcal{D}\left( ^{\mathcal{R}}\Phi \right) \exp \left\{ -%
\mathcal{L}^{(2)}\right\}  \label{146}
\end{equation}%
where%
\begin{equation}
\mathcal{L}^{(2)}=\mathcal{\mathcal{L}}_{\mathcal{R}}+\mathcal{\mathcal{L}_{%
\mathcal{LA}}}  \label{147}
\end{equation}%
with $\mathcal{L}_{\mathcal{R}}$ given by (108) and $\mathcal{\mathcal{L}_{%
\mathcal{LA}}}$ given by (126), the replicon fields verifying the
constraints given by (110), (113), (116) and (119). Performing the
integration over the longitudinal-anomalous and the replicon fields in (146)
considering the replicon constraints, leads to%
\begin{equation}
\left[ \overline{Z^{n}}\right] _{fluct}=\exp \left\{ -\frac{1}{2}\underset{%
r=0}{\sum^{R}}\underset{t=0}{\sum^{R+1}}\mu (t)\ln \hat{\Lambda}%
_{t}^{r}\right. -\frac{1}{2}\underset{t=0}{\sum^{R+1}}\mu (t)\ln \det \hat{%
\Delta}_{t}\qquad  \label{148}
\end{equation}%
\begin{eqnarray*}
&&+\frac{1}{2}\underset{r=0}{\sum^{R}}\left[ \underset{t=0}{\sum^{r}}\mu
(t)\ln \left( 2\left( 1-\frac{p_{r+1}}{p_{r}}\right) \right) \right. +\mu
(r+1)\ln \left( 1-2\frac{p_{r+1}}{p_{r}}\right) \\
&&\qquad \qquad \qquad \qquad \qquad \qquad +\left. \underset{t=r+2}{%
\sum^{R+1}}\mu (t)\ln \left( 1-\frac{p_{r+1}}{p_{r}}\right) \right]
\end{eqnarray*}

\begin{equation*}
-\frac{1}{2}\underset{r=0}{\sum^{R}}\underset{u,v=r+2}{\sum^{R+1}}\mu
(r;u,v)\left( \ln \hat{M}_{u,v}^{r,r}+\frac{1}{\left( \frac{p_{r}}{p_{r+1}}%
-1\right) }\ln \left( \frac{1}{p_{r+1}}\right) -\ln 2\right)
\end{equation*}%
\begin{equation*}
-\frac{1}{2}\underset{r=0}{\sum^{R}}\underset{v=r+2}{\sum^{R+1}}\mu
(r;r+1,v)\left( \ln \hat{M}_{r+1,v}^{r,r}+\frac{1}{\left( \frac{p_{r}}{%
p_{r+1}}-2\right) }\ln \left( \frac{1}{p_{r+1}}-\frac{1}{p_{r}}\right) -\ln
2\right)
\end{equation*}%
\begin{equation*}
-\frac{1}{2}\underset{r=0}{\sum^{R}}\underset{u=r+2}{\sum^{R+1}}\mu
(r;u,r+1)\left( \ln \hat{M}_{u,r+1}^{r,r}+\frac{1}{\left( \frac{p_{r}}{%
p_{r+1}}-2\right) }\ln \left( \frac{1}{p_{r+1}}-\frac{1}{p_{r}}\right) -\ln
2\right)
\end{equation*}%
\begin{equation*}
-\frac{1}{2}\underset{r=0}{\sum^{R}}\mu _{1}(r;r+1,r+1)\left( \ln \hat{M}%
_{r+1,r+1}^{r,r}+\frac{2}{\left( \frac{p_{r}}{p_{r+1}}-3\right) }\ln \left(
2\left( \frac{1}{p_{r+1}}-\frac{1}{p_{r}}\right) \right) \right)
\end{equation*}%
\begin{equation*}
-\frac{1}{2}\underset{r=0}{\sum^{R}}\mu _{2}(r;r+1,r+1)\left( \ln \hat{M}%
_{r+1,r+1}^{r,r}+\frac{2}{\left( \frac{p_{r}}{p_{r+1}}-3\right) }\ln \left( 
\frac{1}{p_{r+1}}-\frac{2}{p_{r}}\right) \right)
\end{equation*}

\noindent where $\hat{\Lambda}_{t}^{r}$ is given by (132) and $\hat{\Delta}%
_{t}$ is%
\begin{equation}
\hat{\Delta}_{t}^{r,s}=\delta _{r,s}^{Kr}+\frac{1}{4}\sqrt{\delta
_{r}^{(t-1)}}\frac{\hat{M}_{t}^{r,s}}{\hat{\Lambda}_{t}^{r}}\sqrt{\delta
_{s}^{(t-1)}}  \label{149}
\end{equation}

\noindent the longitudinal-anomalous multiplicity $\mu (t)$ is given by
(130) and the replicon multiplicities $\mu (r;u,v)$ for the various cases of 
$u,v\geqslant r+1$ are given by (111), (114), (117) and (120).

One observes that in (148) there is a cancellation of terms between the
longitudinal-anomalous and the replicon contributions. Hence, one obtains
for the free-energy fluctuations,

\begin{equation}
F_{fluct}=\frac{1}{\beta }\underset{n\rightarrow 0}{\lim }\frac{1}{2n}%
\left\{ \underset{t=0}{\sum^{R+1}}\mu (t)\ln \det \hat{\Delta}_{t}\right.
\label{150}
\end{equation}%
$\qquad $%
\begin{equation*}
+\underset{r=0}{\sum^{R}}\underset{u,v=r+1}{\sum^{R+1}}\bar{\mu}(r;u,v)\ln 
\hat{M}_{u,v}^{r,r}
\end{equation*}%
\begin{eqnarray*}
&&\qquad -\frac{1}{2}\underset{r=0}{\sum^{R}}p_{0}\left[ \left( p_{r+1}+%
\frac{1}{p_{r+1}}\right) \ln \left( p_{r+1}\right) \right. \\
&&\qquad \quad \qquad +\left. \left. \left( 1-\frac{1}{p_{r+1}}\right)
\left( \frac{p_{r}}{p_{r+1}}+p_{r}-p_{r+1}-3\right) \ln 2\right] \right\}
\end{eqnarray*}%
\qquad

\noindent with%
\begin{equation}
\bar{\mu}(r;u,v)=\frac{1}{2}p_{0}\left( p_{r}-p_{r+1}\right) \left( \frac{1}{%
p_{u}}-\frac{1}{p_{u-1}}\right) \left( \frac{1}{p_{v}}-\frac{1}{p_{v-1}}%
\right) ,\quad u,v>r+1  \label{151}
\end{equation}%
\begin{equation}
\bar{\mu}(r;r+1,v)=\frac{1}{2}p_{0}\left( p_{r}-p_{r+1}\right) \left( \frac{1%
}{p_{v}}-\frac{1}{p_{v-1}}\right) \frac{1}{p_{r+1}},\quad v>r+1  \label{152}
\end{equation}%
\begin{equation}
\bar{\mu}(r;r+1,r+1)=\frac{1}{2}p_{0}\left( p_{r}-p_{r+1}\right) \frac{1}{%
p_{r+1}^{2}}.  \label{153}
\end{equation}

\noindent which are the proper multiplicities as remarked in [19,20] ($\bar{%
\mu}=\mu _{reg}$ in their notation).

For $R=0$, (150) reduces to%
\begin{equation}
F_{fluct}=\frac{1}{\beta }\underset{n\rightarrow 0}{\lim }\frac{1}{2n}%
\left\{ \ln M_{L}+(n-1)\ln M_{A}+\frac{1}{2}n(n-3)\ln M_{R}\right\} .
\label{154}
\end{equation}

A discussion on the fluctuations of the free energy is provided in [22],
where it is concluded that the longitudinal-anomalous contribution vanishes,
the full answer being then given by the replicon contribution.

\bigskip \bigskip

\noindent {\large 6. Conclusion }

\bigskip

We developed a field theory for spin glasses using RFT, for the case of
replica symmetry and the case of replica symmetry breaking on an ultrametric
tree, with the number of replicas \textit{n} and the number of replica
symmetry breaking steps \textit{R} generic integers. We defined a new basis
in terms of the RFT of the two-replica fields which block-diagonalizes the
four-replica mass matrix into the replicon, anomalous and longitudinal
modes. As a result, we have a field theory that is directly defined in terms
of the replicon, anomalous and longitudinal fields, in RFT space. The
corresponding eigenvalues are given in terms of the mass RFT. The
propagators in RFT\ space are obtained by inversion of the block-diagonal
matrix, explicit forms are provided for the propagators, which are
particularly simple in the replicon sector. The formalism allows to express
any \textit{i}-replica vertex in the new basis and hence enables to perform
a standard perturbation expansion. Via a clear sequence of steps one can
transform the interaction vertices of the fluctuations in direct replica
space into the interaction vertices of the replicon, anomalous and
longitudinal modes in RFT space, for higher order calculations in the
perturbation expansion.

In the field theory developed for spin glasses with replica symmetry
breaking in direct replica space [18], the free propagators are given by a
fairly complicated set of coupled integral equations, which were solved in
different momenta regimes. Also, the block-diagonalization and inversion of
the mass-matrix performed in direct replica space [19], using a particular
basis, involves a rather difficult procedure. In [20] a Dyson like equation
related the propapagors to the mass operators. In [21] the
block-diagonalization and inversion of the mass-matrix was achieved by
applying directly the RFT on the four-replica mass-matrix. This is to
contrast with the field theory in RFT space that we present.

The field theory in RFT space provides a new tool to investigate the
behaviour of spin glasses. We applied the formalism to calculate the
contribution of the Gaussian fluctuations around the Parisi mean field
solution for the free energy of an Ising spin glass. We also showed that the
propagators in the direct replica space can be simply related to the
propagators in the RFT space, which enables to calculate important physical
quantities. The Gaussian propagators, in addition of being building blocks
of the interacting theory, also have a direct physical meaning [18]. They
are related to correlation functions that reflect the structure of the phase
space. Various components of the propagator in direct replica space
represent overlaps of spin-spin correlation functions inside a single state
and between different states. Physical observables such as the spin glass
and the nonlinear susceptibilities are expressed in terms of the
propagators, having contributions from both intra- and interstate
correlations. It is important to evaluate the contribution of the different
fluctuation sectors, replicon, anomalous and longitudinal, to the various
quantities. An investigation on spin-spin correlation functions in spin
glasses was performed in [23]. We expect that the RFT field theory will
allow to further study the properties of the glassy phase, and hence
contribute for the understanding of spin glasses.

\bigskip \bigskip

\noindent {\large Acknowledgements}

\medskip

\noindent IRP acknowledges the support from Funda\c{c}\~{a}o para a Ci\^{e}%
ncia e a Tecnologia, UI0618, Portugal.

\bigskip

\noindent {\large References}

\medskip

\noindent [1] Binder K and Young A P 1986 \textit{Rev. Mod. Phys.} \textbf{58%
} 801

\noindent [2] M\'{e}zard M, Parisi G and Virasoro M A 1987 \textit{Spin
Glass Theory and Beyond} (Singapore: World Scientific)

\noindent [3] Young A P (ed) 1998 \textit{Spin Glasses and Random Fields }%
(Singapore: World Scientific)

\noindent [4] De Dominicis C and Giardina I 2006 \textit{Random Fields and
Spin Glasses }(Cambridge: Cambridge University Press)

\noindent [5] Parisi G 1979 \textit{Phys.} \textit{Rev. Lett.} \textbf{43}
1754; Parisi G 1980 \textit{J. Phys. }A \textbf{13} 1101; Parisi G 1980 
\textit{J. Phys. }A \textbf{13} 1887; Parisi G 1983 \textit{Phys. Rev. Lett.}
\textbf{50} 1946

\noindent [6] Sherrington D and Kirkpatrick S 1975 \textit{Phys. Rev. Lett.} 
\textbf{35} 1792

\noindent [7] Fisher D S and Huse D A 1986 \textit{Phys. Rev. Lett.} \textbf{%
56} 1601; 1988 \textit{Phys. Rev. }B \textbf{38} 386

\noindent \lbrack 8] Bray A J and Moore M A 1986 \textit{Heidelberg
Colloquium on Glassy Dynamics }Lecture Notes in Physics vol \textbf{275} ed
J L van Hemmen and I Morgenstern (Berlin: Springer) p 121

\noindent [9] McMillan W L 1984 \textit{J. Phys. }C \textbf{17} 3179

\noindent [10] Edwards S F and Anderson P W 1975 \textit{J. Phys. }F\textit{%
: Metal Phys.} \textbf{5} 965

\noindent [11] de Almeida J R L and Thouless D\ J 1978 \textit{J. Phys. }A 
\textbf{11 }983

\noindent [12] Bray A J and Moore M A 1979 \textit{J. Phys. }C \textbf{12} 79

\noindent [13] Pytte E and Rudnick J 1979 \textit{Phys. Rev. }B \textbf{19}
3603

\noindent [14] Guerra F 2003 \textit{Comm. Math. Phys.} \textbf{233} 1\ 

\noindent [15] De Dominicis C and Kondor I 1983 \textit{Phys. Rev. }B 
\textbf{27} 606

\noindent [16] Kondor I and De Dominicis C 1986 \textit{Europhys. Lett. }%
\textbf{2 }617

\noindent [17] De Dominicis C and Kondor I 1985 \textit{J. Phys. Lett. France%
}\textbf{\ 46} L1037\ \ 

\noindent [18] De Dominicis C, Kondor I and Temesv\'{a}ri T 1998 \textit{%
Spin Glasses and Random Fields }ed A P Young (Singapore: World Scientific) p
119

\noindent [19] Temesv\'{a}ri T, De Dominicis C and Kondor I 1994 \textit{J.
Phys. }A \textbf{27} 7569

\noindent [20] De Dominicis C, Kondor I and Temesv\'{a}ri T 1994 \textit{J.
Physique }I \textbf{4} 1287

\noindent [21] De Dominicis C, Carlucci D M and Temesv\'{a}ri T 1997 \textit{%
J. Phys. }I\textit{\ France} \textbf{7} 105

\noindent [22] De Dominicis C and Di Francesco P 2003 \textit{J. Phys. }A 
\textbf{36} 10955

\noindent [23] De Dominicis C, Giardina I, Marinari E, Martin O C and
Zuliani F 2005 \textit{Phys. Rev.} B \textbf{72} 014443

\end{document}